\documentclass[%
reprint,
showpacs,preprintnumbers,
showkeys,
nofootinbib,
 amsmath,amssymb,
 aps,
prb,
 floatfix,
]{revtex4-1}

\usepackage[pdftex]{graphicx}
\usepackage{multirow}
\usepackage[bookmarks=false]{hyperref}
\usepackage{microtype}

\usepackage{array}
\newcommand{\fmone}{\footnotemark[1]}
\newcommand{\fmtwo}{\footnotemark[2]}
\newcommand{\fmthr}{\footnotemark[3]}
\newcommand{\fmfou}{\footnotemark[4]}
\newcommand{\fmfiv}{\footnotemark[5]}
\newcommand{\fmsix}{\footnotemark[6]}
\newcommand{\fmsev}{\footnotemark[7]}
\newcommand{\fmeig}{\footnotemark[8]}
\newcommand{\fmnin}{\footnotemark[9]}
\newcommand{\fmten}{\footnotemark[10]}
\newcommand{\fmele}{\footnotemark[11]}
\newcommand{\fmtwe}{\footnotemark[12]}
\newcommand{\fmtht}{\footnotemark[13]}
\newcommand{\fmfor}{\footnotemark[14]}
\newcommand{\fmfit}{\footnotemark[15]}
\newcommand{\fmsit}{\footnotemark[16]}
\newcommand{\fmsee}{\footnotemark[17]}
\newcommand{\fmeit}{\footnotemark[18]}
\newcommand{\fmnie}{\footnotemark[19]}
\newcommand{\fmtwn}{\footnotemark[20]}

\newcolumntype{C}{>{\raggedright\arraybackslash}p{0.8in}}
\newcolumntype{H}{>{\raggedright\arraybackslash}p{1.5in}}
\newcolumntype{V}{>{\raggedright\arraybackslash}p{1.2in}}
\begin{document}
\title{MEAM potentials for Al, Si, Mg, Cu, and Fe alloys}

\author{B.~Jelinek}
\author{S.~Groh}
  \altaffiliation[Presently at ]{Institute of Mechanics and Fluid Dynamics,
    TU Bergakademie Freiberg,
    Lampadiusstr. 4, 09596 Freiberg, Germany
  }
\author{M.~F.~Horstemeyer}
  \altaffiliation[Also at ]{
Department of Mechanical Engineering,
    Mississippi State University,
    Mississippi State, MS 39762
  }
\affiliation{Center for Advanced Vehicular Systems,
  200 Research Boulevard,
  Starkville, MS 39759
}

\author{J.~Houze}
\author{S.~G.~Kim}
\altaffiliation[Also at ]{Center for Computational Sciences,
  Mississippi State University,
  Mississippi State, MS 39762
}
\affiliation{Department of Physics and Astronomy,
    Mississippi State University,
    Mississippi State, MS 39762
}

\author{G.~J.~Wagner}
\affiliation{Sandia National Laboratories,
  P.O. Box 969, MS 9401,
  Livermore, CA 94551
}

\author{A.~Moitra}
\affiliation{Department of Chemical Engineering,
  The Pennsylvania State University,
  University Park, PA 16802
}

\author{M.~I.~Baskes}
  \altaffiliation[Also at ]{Los Alamos National Laboratory,
    MST-8, MS G755,
    Los Alamos, NM 87545
  }
\affiliation{Mechanical and Aerospace Engineering,
  University of California in San Diego,
  La Jolla, CA 92093
}

\date{\today}

\begin{abstract}
  
  A set of Modified Embedded Atom Method (MEAM) potentials for the
  interactions between Al, Si, Mg, Cu, and Fe was developed from a
  combination of each element's MEAM potential in order to study
  metal alloying.
  Previously published MEAM parameters of single elements have been
  improved for better agreement to the generalized stacking fault
  energy (GSFE) curves when compared with ab-initio generated GSFE
  curves.
  The MEAM parameters for element pairs were constructed based on the
  structural and elastic properties of element pairs in the NaCl
  reference structure garnered from ab-initio calculations, with
  adjustment to reproduce the ab-initio heat of formation of the most
  stable binary compounds. The new MEAM potentials were validated by
  comparing the formation energies of defects, equilibrium volumes,
  elastic moduli, and heat of formation for several binary compounds
  with ab-initio simulations and experiments.
  Single elements in their ground state crystal structure were
  subjected to heating to test the potentials at elevated
  temperatures. An Al potential was modified to avoid formation of an
  unphysical solid structure at high temperatures. The thermal expansion
  coefficient of a compound with the composition of AA~6061 alloy was
  evaluated and compared with experimental values. 
  MEAM potential tests performed in this work, utilizing the universal
  Atomistic Simulation Environment (ASE), are distributed to
  facilitate reproducibility of the results.

\end{abstract}

\pacs{%
61.50.Lt, 
62.20.D-  
61.72.J-  
68.35.-p  
}

\maketitle

\section{Introduction}

Historically, materials have been developed through the correlation of
processing and properties. Several
implementations of materials science principles have given birth to an
engineering framework for materials design.  Over the past two
decades, more efficient computational methodologies have been
developed and the computational power have increased enormously,
making the computational materials design an essential cost-effective
tool to design materials properties. Since materials complexities can
limit the degree of predictability, several time- and length-scale
methodologies (hence spatiotemporal hierarchy) for computational
materials design naturally evolved (cf.\
\citet{horstemeyer2010:multiscale_mod_review} for a review). Out of
several computational methodologies, atomistic simulations not only
can predict the materials properties from a statistical viewpoint, but
can also quantify the mechanisms of the structure-property
relationship. One of the most critical components of atomistic simulations is the
interatomic potential, which determines the forces on individual
atoms. First-principles calculations certainly are capable of
providing very reliable interatomic potentials in a variety of
chemical environments.  However, realistic simulations of alloy
systems, which are essential to reveal many macroscopic materials
properties, often require a number of atoms that renders these methods
impractical -- they either require too much computer memory or take
too long to be completed in a reasonable amount of time.  One
alternative is to use (semi-)empirical interaction potentials that can
be evaluated efficiently, so that the atomistic approaches that use
them can, in certain cases, handle systems with more than a million
atoms.

The Embedded-Atom Method (EAM) is a widely used atomic level
semiempirical model for metals, covalent materials, and
impurities~\cite{daw83:semiem_hydrog_embrit,*daw84:embed,*baskes87:applic,*daw89:model}.
MEAM (Modified EAM) incorporates angular dependency of electron
density into EAM\@. Atomistic simulations of a wide range of elements
and alloys have been performed using MEAM potentials. MEAM model was
first used for silicon, germanium, and their
alloys~\cite{baskes89:semiem}. It was applied to 26 single
elements~\cite{baskes92:modif} and to
silicon-nickel~\cite{baskes94:atomis} alloys and interfaces.
\citet{gall00:atomis} have used MEAM to model tensile debonding of an
aluminum-silicon interface. \citet{lee00:secon} improved MEAM to
account for the second nearest-neighbor interactions. Also,
\citet{huang95:molec} used MEAM and two other potentials to determine
defect energetics in beta-SiC.  MEAM parameters for
nickel~\cite{baskes97:deter} and molybdenum-silicon
system~\cite{baskes99:atomis} were determined by Baskes. MEAM
potentials for Cu, Ag, Au, Ni, Pd, Pt, Al, and Pb based on the first
and the second nearest-neighbor MEAM were constructed by
\citet{lee03:semiem_cu_ag_au_ni}. \citet{hu2001:analy_hcp,hu03:point_hcp}
proposed a new analytic modified EAM many-body potential and applied
it to 17 hcp metals.  The structural properties of various polytypes
of carbon were described using a MEAM potential by
\citet{lee05:carbon}. Recent work of \citet{lee2010:meam_progr}
summarized available MEAM potentials for single elements and alloys.
Several of
these potentials were then used to perform large scale atomistic
simulations to understand the intriguing nature of the ductile and
brittle fracture~\cite{kang2007:brit_duct_semi_nanow_md},
structure-property relationship~\cite{gates2005:comp_mat_ms_nano},
dislocation
dynamics~\cite{noronha2002:disloc_pin_frac_at_dd,Martinez2008}, and
nature of materials fracture~\cite{potir06:molec, Swygenhoven2004}.

\begin{table*}[!bthp]

  \caption{Set of the MEAM potential parameters for single elements.
    The reference structures for Al, Si, Mg, Cu, and Fe are fcc, diamond,
    hcp, fcc, and bcc, respectively.
    $E_\text{c}$ is the cohesive energy, $a_0$ is the equilibrium lattice
    parameter, $A$ is the scaling factor for the embedding energy,
    $\alpha$ is the exponential decay factor for the universal energy,
    $\beta^{(0-3)}$ are the exponential decay factors for the atomic
    densities, $t^{(0-3)}$ are the weighting factors for the atomic
    densities, $C_{\text{max}}$ and $C_{\text{min}}$ are screening
    parameters, $\rho_{0}$ is the density scaling factor that is relevant only
    for element pairs. Definition of these parameters may be found
    in Ref.~\onlinecite{baskes92:modif}.
    Non-zero parameters $\delta_r$ in Rose Eq.~(\ref{eq:rose1}--\ref{eq:rose4})
    were used for Al ($\delta_r=0.1$) and Fe ($\delta_r=0.3$),
    along with $\delta_a=0.0$.
    \label{tab:meam_pars}}
  \begin{ruledtabular}
    \begin{tabular}{ccccccccccccccccc}
      elem. &
      $E_\text{c}$[eV] & $a_0$[\AA] & $A$ & $\alpha$ &
      $\beta^{(0)}$ & $\beta^{(1)}$ & $\beta^{(2)}$ & $\beta^{(3)}$ &
      $t^{(0)}$ & $t^{(1)}$ & $t^{(2)}$ & $t^{(3)}$ &
      $C_{\text{min}}$ & $C_{\text{max}}$ & $\rho_0$ \\
      \hline
      Al    &
      3.353 & 4.05 &  1.07 & 4.64 &
      2.04  & 3.0 &  6.0  & 1.5  &
      1.0   & 4.50 & -2.30 & 8.01 &
      0.8   & 2.8  & 1.0 \\
      Si    &
      4.63  & 5.431&  1.00 & 4.87 &
      4.4   & 5.5  &  5.5  & 5.5  &
      1.0   & 2.05 &  4.47 & -1.80&
      2.0   & 2.8  & 2.2 \\
      Mg    &
      1.51  & 3.194& 0.8  & 5.52 &
      4.0   & 3.0  & 0.2  & 1.2  &
      1.0   & 10.04& 9.49 & -4.3 &
      0.8   & 2.8  & 0.63 \\
      Cu    &
      3.54  & 3.62 &  1.07 & 5.11 &
      3.634 & 2.2 &  6.0  & 2.2 &
      1.0   & 4.91 & 2.49 & 2.95  &
      0.8   & 2.8  & 1.1 \\
      Fe    &
      4.28  & 2.851& 0.555 & 5.027&
      3.5   & 2.0 &  1.0  & 1.0 &
      1.0   & -1.6 & 12.5 & -1.4  &
      0.68  & 1.9  & 1.0 \\
    \end{tabular}
  \end{ruledtabular}
\end{table*}

Aluminum, magnesium, copper, and iron alloys are being used in
developing materials with novel properties. Great popularity of these
alloys is connected to their general functional properties, mechanical
properties, mass density, corrosion resistance, and machinability.
Light metal alloys, such as magnesium and aluminum alloys, are now
demanded for use in the automotive and aviation industries.  They
performed remarkably well for the purpose of decreasing the operating
expenses and fuel consumption. These alloys usually contain several
other minor elements, such as silicon, nickel, and manganese, and are
known to have very complex phase compositions.  Assessment of such
complex systems is a very challenging task, since different
constituent elements can form different phases, whose selection
depends on the ratio between the constituents and also on a variety of
processing and treatment factors.

Contrary to DFT potentials, most of the single element semiempirical
potentials do not combine easily into multi-component alloy models.
The difficulty of combining single element EAM potentials into alloy
systems comes from the need of their
normalization~\cite{johnson1989:analyt_eam_bcc}. The procedure to form
EAM alloy parameterization from single element potentials was
suggested~\cite{johnson1989:alloy,zhou2001:atom_multil}, but it does
not guarantee that the resulting potential will be suitable for
modeling compounds~\cite{zhou04:misfit}. Alloy potentials usually
introduce new parameters for each pair of elements, allowing
to fit properties of their binary compounds. The number of
parameters to adjust and the number of tests to perform is
proportional to the square of the number of constituent elements.
In the present MEAM approach, each pair interaction is characterized
by a total of 13 parameters (Table~\ref{tab:meam_parp}, and the ratio
of density scaling factors $\rho_0$ for constituent elements,
Table~\ref{tab:meam_pars}). Adoption of the default value
C$_\text{max}=2.8$ leads to 9 adjustable parameters for each
pair. Comparable angularly dependent potentials for the Fe-Ni system
\cite{mishin2005:phase_Fe-Ni} also have 9 adjustable pair parameters.

While the semiempirical potentials have been developed and tested for
binary alloys
\cite{mendelev09:dev_pot_s-l_almg,Liu19983467,apostol2011:interat_alcu,liu1999:new_alcu,mendelev2005:effect_fe_al}, 
binaries, similarly to single element potentials, may not combine
easily into ternaries. Modeling of ternary systems faces a challenge
since less experimental properties are available for ternary systems.
Ternary potentials are usually examined only at a particular composition
range---the number of possible compositions grows to the power of the
number of constituent elements. It is also nontrivial to find an
equilibrium structure for complex systems of representative size at
low temperatures.  Ternary potentials are only available for
Fe/Ni-Cr-O~\cite{ohira2001:magnetite_meam,*ohira1997:atom_adh_metalox}
(MEAM), Pu-Ga-He~\cite{valone2006:at_he_bubbles} (MEAM), Fe-Ti-C/N,
Cu-Zr-Ag, Ga-In-N,
Fe-Nb-C/N~\cite{kim2009:meam_fe-ti-c-n,kang2009:cu-zr-ag,do2009:at_35_nitr,kim2010:meam_fenbcn}
(MEAM), H-C-O~\cite{chenoweth2008:reaxff_hco} (Reactive Force Field,
ReaxFF),
Ni-Al-H~\cite{angelo1995:trap_h_ni,*baskes1997:trap_h_ni_corr},
Zr-Cu-Al~\cite{cheng:atom_glass_eam_zr-cu-al}, and
Fe-Cu-Ni~\cite{bonny2009:tern_fecuni} (EAM) systems.
To extend from binaries to ternaries, MEAM provides a ternary screening
parameter C$^\text{XYZ}$. In the present work we did not examine
ternary systems. Instead, we performed thermal expansion simulations
of a compound including all species of the potential. The default
values of $C_\text{min}=2.0$ and $C_\text{max}=2.8$ were applied for
ternary screening. Since an effort beyond the scope of our project is
required for satisfiable validation of the 5-element alloy
potential under varying temperatures, compositions, and configuration states,
we concentrated on basic tests and on providing tools to facilitate
reproducibility of the tests~\cite{site:ase_tests}.

In the present study we develop a MEAM potential for aluminum,
silicon, magnesium, copper, iron, and their combinations.  We fit the
potential to the properties of single elements and element pairs, but
the model implicitly allows calculations with any combination of
elements. We show the applied MEAM methodology in
Appendix~\ref{sec:meam_theory}. The DFT calculations are described in
Sec.~\ref{app:ab-initio}.  In Sec.~\ref{sec:single}, the
single-element volume-energy curves in basic crystal structures, and
also important material properties, such as formation energies of
vacancies, self-interstitials, surfaces, and generalized stacking
fault energies from MEAM are examined and compared with DFT
calculations. In Sec.~\ref{sec:pairs}, the MEAM potential parameters
for each unlike element pair are initialized to fit the ab-initio heat
of formation, equilibrium volume, and elastic moduli of the
hypothetical NaCl reference structure. Heat of formation of binary
compounds in a variety of crystal structures from MEAM are thereafter
examined and compared with the ab-initio and experimental results. The
MEAM parameters are adjusted to match the DFT formation energy of the
most stable compounds. The structural and elastic properties for
several binary compounds and formation energies of substitutional
defects are compared with ab-initio and experimental results. Finally
we performed thermal expansion simulations of a compound with the
composition of an AA~6061 alloy (\ref{sec:finite_t}). We conclude with a
short summary.

\section{Ab-initio calculations}
\label{app:ab-initio}

Ab-initio total energy calculations in this work were based on density
functional theory (DFT), using the projector augmented-wave (PAW)
method~\cite{blochl1994:paw,*kresse1999:from_us_paw} as implemented in
the {\sc VASP} code~\cite{kresse93:abinitio,*Kresse:1996:VASP:PRB-69}.
Exchange-correlation effects were treated by the generalized gradient
approximation (GGA) as parameterized by
\citet{perdew1992:pw91}.  
All DFT calculations were performed in high precision with the
plane-wave cut-off energy set to 400~eV in order to achieve the
convergence of heat of formation and elastic properties. Integration
over the irreducible Brillouin zone was performed using the
$\Gamma$-centered Monkhorst-Pack scheme~\cite{monkhorst76:mp} with the
size gradually increased to 7$\times$7$\times$7 for point defects, to
19$\times$19$\times$1 for surfaces, and to 29$\times$29$\times$29 to
improve convergence of shear moduli at small strains. Elastic
constants presented here were obtained without relaxation of atomic
positions. Since most of the examined high energy structures are, at
best, metastable, relaxation does not maintain the crystal symmetry,
resulting in large energy changes and unphysical elastic constants.

\section{MEAM parameters for single elements}
\label{sec:single}

The present MEAM parameters for single elements are listed in
Table~\ref{tab:meam_pars}. The initial values of these parameters were
taken from existing MEAM
potentials~\cite{baskes92:modif,lee03:semiem_cu_ag_au_ni,jelinek2007:meam_mgal,lee2011:meam_fe}.
The $C_{\text{min}}$ screening parameter for Al,
Mg, and Cu was lowered from 2.0 to 0.8 to improve the GSFE curves
(Sec.~\ref{sec:sf}).
The Mg potential was
adjusted to reproduce the DFT values of hcp, bcc, and fcc energy
differences, vacancy formation energy, and $(10\bar{1}0)$ surface
formation energy.
The Al potential was modified to prevent formation
of an unknown structure at elevated temperatures (Sec.~\ref{sec:finite_t}).

\subsection{Energy dependence on volume of single elements in fcc,
  hcp, bcc, and simple cubic crystal structures}

\begin{figure}[!htbp]
  \includegraphics[]{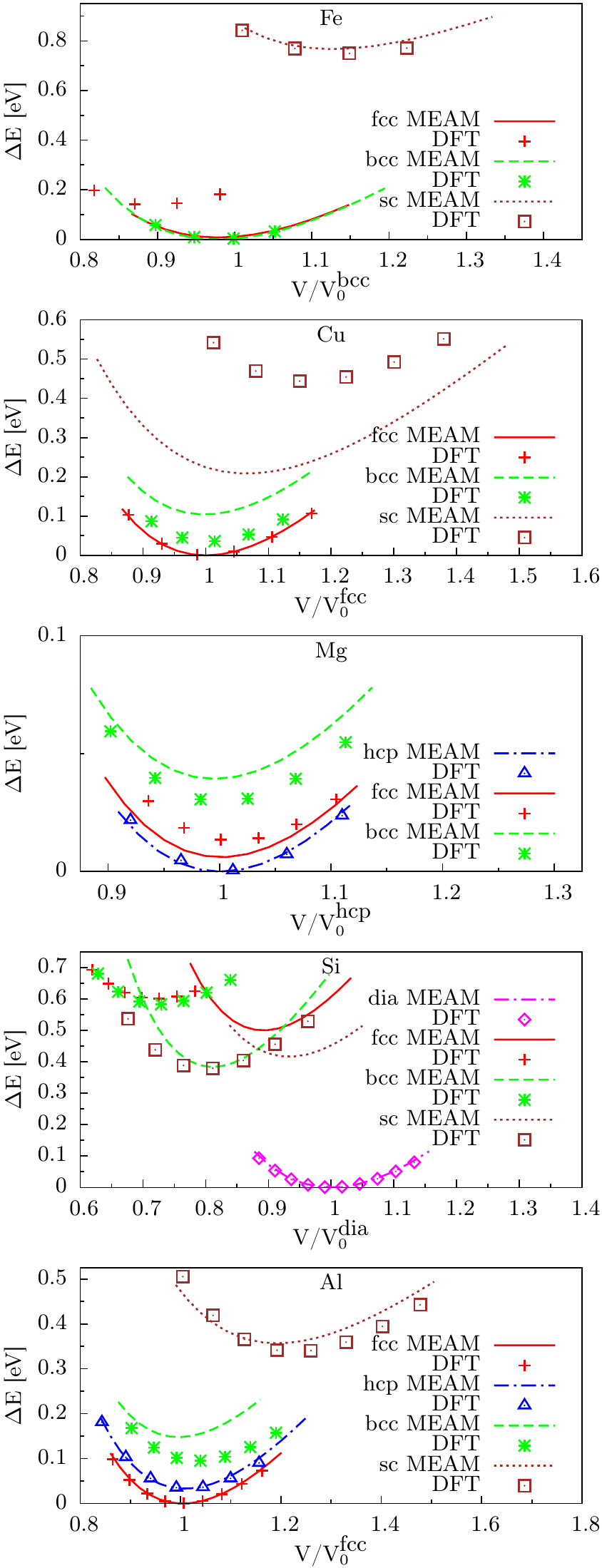}
  \caption{(Color online) Energy-volume dependence of Al, Si, Mg, Cu,
    and Fe in fcc, hcp, bcc, diamond, and simple cubic crystal structures
    relative to the ground state.}
  \label{fig:ve}
\end{figure}

The first test of the validity of MEAM potential for single elements
is a comparison of the energy-volume curves in the fcc, hcp, bcc,
diamond, and simple cubic crystal structures, shown in
Fig.~\ref{fig:ve}. The MEAM potentials appropriately capture the
lowest energy structures of Al (fcc), Si (dia), Mg (hcp), Cu (fcc),
and Fe (bcc). Also, the equilibrium volumes of several crystal
structures from MEAM closely match the DFT results. Better match of
DFT energy differences and volume ratios can possibly be obtained by
optimization of Si and Cu MEAM parameters. Fe MEAM potential applied
in the present work is a MEAM-p variant of Fe potential from the
recent effort of \citet{lee2011:meam_fe}, exhibiting a correct low
temperature phase stability with respect to the pressure. The fcc
equilibrium energy and volume from this Fe potential is very close to
the bcc equilibrium in order for the structural transition to appear
at finite temperature without magnetic contribution.
In general the MEAM potentials of the present work reproduced the DFT
results for the individual elements fairly well.

\subsection{Vacancies}

The formation energy of a single vacancy $E_{\text{f}}^{\text{vac}}$
is defined as the energy cost to create a vacancy:
\begin{equation}
  E_{\text{f}}^{\text{vac}} = E_{\text{tot}}[N] - N\varepsilon,
  \label{eq:Ef_v}
\end{equation}
where $E_\text{tot}[N]$ is the total relaxed energy of a system with
$N$ atoms containing a vacancy and $\varepsilon$ is the energy per
atom in the bulk. Cell volume and atomic positions were relaxed in
each case. Table~\ref{tab:vac} shows the formation energies of a
single vacancy for the fcc Al cell, diamond Si cell, hcp Mg, fcc Cu,
and bcc Fe obtained from the MEAM and DFT calculations.  The MEAM
systems sizes were 5$\times$5$\times$5 primitive fcc and bcc cells,
3$\times$3$\times$3 primitive diamond cells, and 8$\times$4$\times$4
orthogonal hcp cells. For the DFT systems, the simulation sizes were
5$\times$5$\times$5 fcc, 4$\times$4$\times$4 diamond and bcc,
and 4$\times$4$\times$2 hcp primitive cells.

The vacancy formation energy of Mg was slightly improved in comparison
with previous MEAM results~\cite{jelinek2007:meam_mgal}. Overall
agreement of vacancy formation energies between MEAM, experiment,
and DFT was within a few eV, and the present results are comparable
or better than those from other calculations. The reduction in
volume due to the formation of a vacancy agrees well with the DFT, except
the value for Fe is somewhat low.

\begingroup
\squeezetable
\begin{table}[!tbhp]
  \caption{Calculated single vacancy properties. 
    Single vacancy formation energy $E_{\text{f}}^{\text{vac}}$ and
    formation volume $\Omega_v$ values are obtained from the relaxed
    structures containing single vacancies. Here $\Omega_0$ is the
    bulk atomic volume. All energy values are 
    listed in eV.  The results from the MEAM calculations
    are compared with the results from the DFT calculations given inside
    the parentheses, other simulations, and experiments.
    \label{tab:vac}}
  \begin{ruledtabular}
    \begin{tabular}{cccccc}
      \multirow{2}{*}{}
         & \multicolumn{2}{c}{$E_{\text{f}}^{\text{vac}}$}
                                                            & \multicolumn{3}{c}{$\Omega_v/\Omega_0$}                  \\
                                                             \cline{2-4} \cline{5-6}
         & Present    & Others                              & Exp        & Present         & Others             \\
                                                             \cline{1-4} \cline{5-6}
      Al & 0.67 (0.5) & 0.68\fmone 0.68\fmtwo (0.55\fmsix)  & 0.67\fmsev & 0.67 (0.7)      & 0.72\fmone 0.61\fmtwo \\
      Si & 3.27 (3.6) & 3.56\fmtwn 3.67\fmnie (3.6\fmthr)   & 3.6\fmnin  & 0.21 (0.3) & 0.94\fmtwn \\
      Mg & 0.89 (0.7) & 0.59\fmfou 0.87\fmtwo (0.83\fmele)  & 0.79\fmfiv & 0.72 (0.8) & 0.83\fmfou 0.87\fmtwo \\
      Cu & 1.10 (1.0) & 1.05\fmsee 1.27\fmeit (1.03\fmeig)             & 1.2\fmten   &0.75 ($\sim$0.9)& 0.70\fmeit 0.74\fmeit \\
      Fe & 1.65 (2.1) & 1.84\fmtwe 1.89\fmtht (1.95\fmfor) & 1.53\fmfit & 0.47 ($\sim$0.8) & 0.6\fmsit (0.90\fmfor)\\
    \end{tabular}
  \end{ruledtabular}
  \footnotetext[1]{MEAM results by~\citet{lee03:semiem_cu_ag_au_ni}}
  \footnotetext[2]{Calculated using EAM parameters extracted from~\citet{Liu:SURF-v373-1997}}
  \footnotetext[3]{DFT calculation by~\citet{wright2006:dft_si_vac}}
  \footnotetext[4]{AMEAM results by~\citet{hu2001:analy_hcp}}
  \footnotetext[5]{Experimental results by~\citet{tzanetakis1976:form_en_vac_al_mg}}
  \footnotetext[6]{DFT results by~\citet{carling2003:vac_conc_al_fp_mp}}
  \footnotetext[7]{Experimental value by~\citet{hehenkamp1994:abs_vac_conc}}
  \footnotetext[8]{DFT calculation by~\citet{andersson2004:mono_div_cu_fp}}
  \footnotetext[9]{Experimental value by~\citet{dannefaer1986:monov_si,*throwe1989:search_monov_si}}
  \footnotetext[10]{Experimental value by~\citet{hehenkamp1992:eq_vac_cu}}
  \footnotetext[11]{DFT value by~\citet{krimmel2000:ab_initio_vac_mg}}
  \footnotetext[12]{EAM value by~\citet{mendelev2003:dev_liq_fe}}
  \footnotetext[13]{Finnis-Sinclair potential value by~\citet{ackland97:comp_fecu}}
  \footnotetext[14]{DFT value by~\citet{domain01:ab_def_fe_fecu}}
  \footnotetext[15]{Experimental value by~\citet{schaefer1977:vac_fe_positron}}
  \footnotetext[16]{Experimental value referenced in~\citet{ackland97:comp_fecu}}
  \footnotetext[17]{EAM value by~\citet{mendelev2008:anal_liq_alcu}}
  \footnotetext[18]{EAM value by~\citet{mishin2001:str_cu_ab_eam}}
  \footnotetext[19]{MEAM value by~\citet{timonova2007:sputterSi}}
  \footnotetext[20]{MEAM value by~\citet{ryu2009:impr_meam_si}}
\end{table}
\endgroup

\subsection{Self-interstitials}

\begin{table}[!t]
  \caption{The formation energies of various
    Al, Si, Mg, Cu, and Fe self-interstitials. All energy
    values are given in eV.  The results from the MEAM calculations
    are compared with the DFT results and other classical MD (CMD) simulations.\label{tab:interstitials}}
  \begin{ruledtabular}
    \begin{tabular}{lcccccc}
    Interstitial &DFT       &MEAM &\multicolumn{3}{c}{CMD} & DFT \\
      \hline                      
    Al&\multicolumn{2}{c}{Present}&    Ref.\cite{Liu:SURF-v373-1997} & Ref.\cite{deDebiaggi2006:theo_selfint_al_ni} & Ref.\cite{PurjaPun2009:md_self_diff_disl_al} &Ref.\cite{sandberg2002:self_diff_al}\\
      tetrahedral&3.3       &3.32 &    3.16\fmone & 2.94 \\
      octahedral &2.8       &3.26 &    3.06\fmone & 2.82 \\
      split (100)&2.7       &2.77 &    2.68\fmone & 2.46 & 2.59 & 2.43 \\
      \hline                      
    Si&\multicolumn{2}{c}{Present}&Ref.\cite{gillespie2007:bond_order_si}& Ref.\cite{tersoff1988:empir_si_elast} & Ref.\cite{erhart2005:anal_pot_si_c} &Ref.\cite{al-mushadani2003:free_def_si}\\
      split (110)&3.7       &3.71 &3.88&4.7 & 3.9 &3.40 \\
      \hline                      
    Mg&\multicolumn{2}{c}{Present}&Ref.\cite{Liu:SURF-v373-1997}      & Ref.\cite{jelinek2007:meam_mgal} & Ref.\cite{kim2009:semi_constr} & Ref.\cite{jelinek2007:meam_mgal}\\
      tetrahedral&2.2       &1.63 &1.53\fmone                         & 1.53 & 1.41 &2.35\\
      octahedral &2.2       &1.57 &2.16\fmone                         & 1.29 & 1.20 &2.36\\
     split (0001)&2.3       &1.78 &1.52\fmone                         &      &      &\\
      \hline                      
    Cu&\multicolumn{2}{c}{Present}&Ref.\cite{mendelev2008:anal_liq_alcu} & Ref.\cite{mishin2001:str_cu_ab_eam} & Ref.\cite{mishin2001:str_cu_ab_eam} & Ref.\cite{mendelev2008:anal_liq_alcu}\\
      tetrahedral&3.9       &3.37 &2.99\fmone&  \\
      octahedral &3.5       &2.72 &2.97\fmone&  \\
      split (100)&3.3       &2.59 &2.81      & 3.06 & 3.23 & 2.93 \\
      \hline                      
    Fe&\multicolumn{2}{c}{Present}&Ref.\cite{mend03:dev_fe}& Ref.\cite{dudarev2005:mag_pot_md} & Ref.\cite{dudarev2005:mag_pot_md} & Ref.\cite{fu2004:stab_interst_fe}\\
      tetrahedral&4.2       &4.31 &4.16\fmone              &            &       & 4.14\\
      octahedral &5.0       &4.78 &4.19\fmone              &            &       & 4.82\\
      split (110)&3.9       &3.79 &3.53                    & 4.11       & 3.65  & 3.64\\ 
      split (111)&4.9       &4.28 &4.02                    & 4.01       & 4.24  & 4.34\\ 
      split (100)&4.8       &4.81 &4.34                    & 4.28       & 4.60  & 4.64\\                
    \end{tabular}
  \end{ruledtabular}
  \footnotetext[1]{Calculated using parameters from the given reference.}
\end{table}

The formation energy of an interstitial point defect
$E_{\text{f}}^{\text{int}}$ is given by
\begin{equation}
  E_{\text{f}}^{\text{int}}
  =
  E_{\text{tot}}[N+1]
  - N \varepsilon_{\text{X}}
  - \varepsilon_{\text{Y}}
  \label{eq:Ef_int}
\end{equation}
where $E_\text{tot}[N+1]$ is the total energy of a system with $N$
type-X bulk atoms plus one impurity atom of type-Y inserted at one of
the interstitial sites, and $\varepsilon_{\text{X}}$
($\varepsilon_{\text{Y}}$) is the total energy per atom of type-X
(type-Y) in its most stable bulk structure. The inserted atom
$\text{Y}$ can be of the same type as the bulk, in which case the
point defect is called a self-interstitial defect.  Self-interstitial
formation energies were calculated for Al, Si, Mg, and Cu at the
octahedral, tetrahedral, and dumbbell sites.  Dumbbell orientations
were [100] for fcc, [0001] for hcp, and [110] for bcc and diamond
structures. Relaxations of the atomic positions and the volume were also
performed, and the DFT and MEAM results are listed in
Table~\ref{tab:interstitials}. Similar to the previous calculations,
the MEAM systems sizes were
5$\times$5$\times$5 primitive fcc and bcc cells, 3$\times$3$\times$3
primitive diamond cells, and 8$\times$4$\times$4 orthogonal hcp
cells. For the DFT systems, the examined sizes were
5$\times$5$\times$5 fcc primitive cells, 4$\times$4$\times$4 diamond
and bcc primitive cells, and 4$\times$4$\times$2 hcp primitive cells.

In general, the DFT results are well reproduced or slightly underestimated
by the MEAM potentials.
According to the present MEAM potential, the most stable form of a
self-interstitial defect for fcc Al is a dumbbell along the
[100] direction, in agreement with the DFT results and an experimental
observation by \citet{jesson1997:thermal_int_al}. The results for Mg
are better than those published
previously~\cite{jelinek2007:meam_mgal,kim2009:semi_constr}. The
present Mg potential indicates that the tetrahedral site will be most
stable in agreement with the DFT calculations. For both Cu and Fe, the
new MEAM potential produces the same relative stability of the examined
interstitial sites with the DFT calculations.

\subsection{Surfaces}

\begin{table}[]
  \caption{Surface formation energies for Al,
    Si, Mg, Cu, and Fe. The units are mJ/$\text{m}^2$. The second column
    indicates if the structure was relaxed. Comparisons with other classical MD (CMD),
    DFT, and experimental values for polycrystalline surfaces and Si facets
    are also given.\label{tab:E_surf}}
  \begin{ruledtabular}
    \begin{tabular}{cccccccc}
      Surface & Rlx & DFT & MEAM & \multicolumn{2}{c}{CMD} & DFT & EXP\\
      \hline
      Al & &\multicolumn{2}{c}{Present}
      & Ref.~\cite{lee03:semiem_cu_ag_au_ni} 
      & Ref.~\cite{Liu:SURF-v373-1997}\footnotemark[2] 
      & Ref.~\cite{vitos1998:the_surf_en_met}
      & Ref.~\cite{deboer1988:coh_tma}
      \\
      (111) & No  &  780 & 820  &     & 913 & 1199 & 1143\\
      (111) & Yes &  780 & 752  & 629 & 912 &\\
      (110) & No  &  990 & 1154 &     & 1113& 1271 \\
      (110) & Yes &  960 & 1135 & 948 & 1107 \\
      (100) & No  &  890 & 1121 &     & 1012& 1347\\
      (100) & Yes &  890 & 1088 & 848 & 1002 \\
      \hline
      Si & &\multicolumn{2}{c}{Present}
      & Ref.~\cite{wilson1990:mod_si_surf_comp}
      & Ref.~\cite{timonova2011:opt_si}
      & Ref.~\cite{stek2002:absol_surf_IV}
      & Ref.~\cite{eaglesham98:eq_si}
      \\
      (111) & No  & 1620 & 1254 & 1405 &      & 1820 &     \\
      (111) & Yes & 1570 & 1196 & 1405 &      & 1740 & 1230\\
      (100) & No  & 2140 & 1850 & 2434 &      & 2390 &     \\
      (100) & Yes & 2140 & 1743 & 1489 &      & 2390 & \\
(100) & 2$\times$1&      & 1241 &      & 2050 & 1450 & 1360 \\
      \hline                        
      Mg & Rlx &\multicolumn{2}{c}{Present} 
      & Ref.~\cite{hu2001:analy_hcp} 
      & Ref.~\cite{Liu:SURF-v373-1997}\footnotemark[2] 
      & Ref.~\cite{vitos1998:the_surf_en_met}
      & Ref.~\cite{deboer1988:coh_tma}
      \\
      (0001)& No  &  530 & 780  &     & 500 & 792 & 785   \\
      (0001)& Yes &  530 & 713  & 310 & 499 &     & \\
(10$\bar{1}$0)&No &  850 & 878  &     & 629 & 782 &    \\
(10$\bar{1}$0)&Yes&  850 & 859  & 316 & 618 \\
      \hline
      Cu & &\multicolumn{2}{c}{Present} 
      & Ref.~\cite{Liu19993227}\footnotemark[2] 
      & Ref.~\cite{mendelev2008:anal_liq_alcu}\footnotemark[2]
      & Ref.~\cite{vitos1998:the_surf_en_met}
      & Ref.~\cite{deboer1988:coh_tma}
      \\
      (111) & No  & 1290 & 1411 & 1185 & 919 &1952& 1825\\
      (111) & Yes & 1290 & 1411 & 1181 & 903 &      \\
      (110) & No  & 1550 & 1645 & 1427 & 1177&2237\\
      (110) & Yes & 1510 & 1614 & 1412 & 1153\\
      (100) & No  & 1440 & 1654 & 1291 & 1097&2166\\
      (100) & Yes & 1430 & 1653 & 1288 & 1083\\
      \hline
      Fe & &\multicolumn{2}{c}{Present}
      & Ref.~\cite{zhou04:misfit}\footnotemark[2] 
      & Ref.~\cite{mend03:dev_fe}\footnotemark[2]
      & Ref.~\cite{blonski2007:str_bcc_fe}
      & Ref.~\cite{deboer1988:coh_tma}\\
      (111) & No  & 2760 & 1366 & 1941 & 2012 & 2660& 2475\\
      (111) & Yes & 2700 & 1306 & 1863 & 1998 & 2580\\
      (110) & No  & 2420 & 1378 & 1434 & 1651 & 2380\\
      (110) & Yes & 2420 & 1372 & 1429 & 1651 & 2370 \\
      (100) & No  & 2500 & 1233 & 1703 & 1790 & 2480\\
      (100) & Yes & 2480 & 1222 & 1690 & 1785 & 2470\\
    \end{tabular}
  \end{ruledtabular}
  \footnotetext[1]{Value from the given reference.}
  \footnotetext[2]{Calculated using parameters from the given reference.}
\end{table}

A semi-infinite surface is one of the simplest forms of defects. To
test the transferability of the new MEAM potentials, formation
energies for several 1$\times$1$\times$7 surface slabs with 8
$\text{\AA}$ vacuum layer were computed. Eight atomic layers were used
for the Si(111) surface and 12 layers for the 2$\times$1 Si(100)
surface reconstruction.
The surface formation energy
per unit surface area $E_{\text{surf}}$ is defined as
\begin{equation}
  \label{eq:surf}
  E^{\text{surf}}_{\text{f}} = 
  \frac
  {E_{\text{tot}}[N] - N \varepsilon
  }{A
  },
\end{equation}
where $E_{\text{tot}}[N]$ is the total energy of the structure with
two surfaces, $N$ is the number of atoms in the structure,
$\varepsilon$ is the total energy per atom in the bulk, and $A$ is the
total area of both surfaces.
Table~\ref{tab:E_surf} shows the surface formation
energies of several surfaces constructed from fcc Al, hcp Mg, fcc Cu,
and bcc Fe crystals.  Results from the present MEAM potentials agree,
in the order of magnitude, with the DFT calculations, except for Fe
values being underestimated.

The surfaces with lowest energy without reconstruction are identified
correctly by the present MEAM potentials. The 2$\times$1
reconstruction of the Si(100) surface leads to symmetric dimers in accord
with other Si potentials~\cite{balamane1992:comp_si_empir}. Note that
surface formation energies from the present PAW GGA calculations
are lower than our previously published
results~\cite{jelinek2007:meam_mgal} using ultrasoft pseudopotentials
within local density (LDA) approximation---it is known that GGA leads
to surface energies which are 7--16\% lower than LDA values for
jellium and 16--29\% lower than the experimental
results\cite{vitos1998:the_surf_en_met,perdew1992:pw91}. A procedure
\cite{mattsson2001:en_funct_surf} and new DFT functionals
\cite{armiento2005:AM05,perdew2008:PBEsol} were suggested to correct
the errors of LDA and GGA approximations. Similar correction can be
applied to vacancy formation energies
\cite{mattsson2006:noneq_pbe_pw91}, but such corrections were not
applied in the present study.

\subsection{Stacking faults}
\label{sec:sf}

Using an assumption of a planar dislocation core, the Peierls-Nabarro
model\cite{Peierls_1940, Nabarro_1947} is a powerful theory to
quantify the dislocation core properties. In that model, a dislocation
is defined by a continuous distribution of shear along the glide
plane, and the restoring force acting between atoms on either sides of
the interface is balanced by the resultant stress of the
distribution. As shown in the recent study of \citet{Carrez_09}, a
solution of the Peierls-Nabarro model can be obtained numerically by
identifying the restoring force to the gradient of the generalized
stacking fault energy (GSFE) curve \cite{Vitek_66}. In addition,
\citet{Swygenhoven2004} claimed that the nature of slip in
nanocrystalline metals cannot be described in terms of an absolute
value of the stacking fault energy, and a correct interpretation
requires the GSFE curve, which shows the change in energy per unit
area of the crystal as a function of the displacement varied on the
slip plane. However, the GSFE curve is not experimentally accessible.
Therefore, to model dislocation properties reliably, the GSFE curve
calculated with the MEAM potential must reproduce the DFT data.

The stacking fault energy per unit area of a stacking fault
$E^{\text{sf}}_{\text{f}}$ is defined as
\begin{equation}
  \label{eq:sf}
  E^{\text{sf}}_{\text{f}} = 
  \frac
  {E_{\text{tot}}[N] - N \varepsilon
  }{A
  },
\end{equation}
where $E_{\text{tot}}[N]$ is the total energy of the structure with
a stacking fault, $N$ is the number of atoms in the structure,
$\varepsilon$ is the total energy per atom in the bulk, and $A$ is the
total area of surface.

\begin{figure*}[]
  \includegraphics []{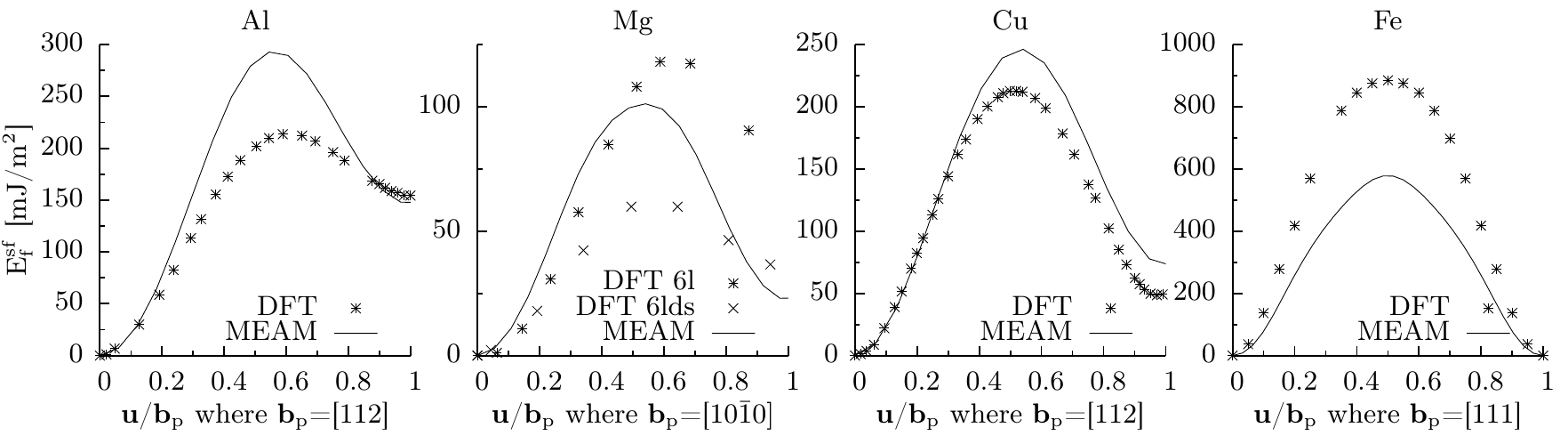} \\
  \caption{GSFE curves for Al, Mg, Cu, and Fe obtained with the MEAM
    potential and compared with the DFT data.}
 \label{fig:gsf}
\end{figure*}

As a validation test of the MEAM potential, the GSFE curves obtained
by molecular statics (MS) were compared with the DFT data by
\citet{Zimmerman_00} for Al and Cu, by the present authors for Fe,
and by \citet{Datta_08} for Mg. After lowering the C$_\text{min}$
parameter to 0.8, the GSFE curves calculated by MS using the MEAM
potential for Al, Cu, and Mg show the skewed sinusoidal shape
in agreement with the DFT predictions (Fig.~\ref{fig:gsf}) illustrating
reasonable agreement with the DFT GSFE curves.

\begin{table*}[!tbhp]
  \caption{The MEAM potential parameters for element pairs.
    $\triangle H_{\text{B1}}^{\text{XY}}$ is the heat of formation
    of the NaCl
    structure (reference) with the type-X and type-Y elements
    relative to the energies of elemental X and Y in their equilibrium
    reference state, $r_e$ is their equilibrium nearest
    neighbor distance, $\alpha$ is the exponential decay factor for the
    universal energy, $C_{\text{max}}$ and $C_{\text{min}}$ are
    screening parameters
    ($C^\text{XYX}$ denotes type-Y element between two type-X elements).
    Non-zero parameters $\delta_r=\delta_a=0.1$
    in Rose Eq.~(\ref{eq:rose1}--\ref{eq:rose4}) were used for SiFe pair.
    \label{tab:meam_parp}}
  \begin{ruledtabular}
    \begin{tabular*}{0.75\textwidth}{cccccrlrlrlrlcc}
      X & Y & $\triangle H_{\text{B1}}^{\text{XY}}$[eV] & $r_\text{e}^{\text{XY}}$[\AA] & $\alpha^{\text{XY}}$ &
      $C_\text{min}^\text{XYX}$ & $C_\text{max}^\text{XYX}$ & 
      $C_\text{min}^\text{YXY}$ & $C_\text{max}^\text{YXY}$ &
      $C_\text{min}^\text{XXY}$ & $C_\text{max}^\text{XXY}$ &
      $C_\text{min}^\text{XYY}$ & $C_\text{max}^\text{XYY}$ \\
      \hline
      Al & Si &  0.28 & 2.62 & 4.56 & 0.5 & 2.8  &  2.0 & 2.8  &  2.0 & 2.8  & 2.0 & 2.8 \\
      Al & Mg &  0.23 & 2.87 & 4.52 & 2.0 & 2.8  &  0.0 & 2.8  &  2.0 & 2.8  & 0.0 & 2.8 \\
      Al & Cu &  0.19 & 2.53 & 4.65 & 0.0 & 2.8  &  2.0 & 2.8  &  2.0 & 2.8  & 2.0 & 2.8 \\
      Al & Fe &  0.26 & 2.45 & 4.64 & 0.9 & 2.8  &  0.1 & 2.8  &  2.0 & 2.8  & 2.0 & 2.8 \\
      Si & Mg &  0.20 & 2.75 & 4.73 & 1.0 & 2.8  &  1.0 & 2.8  &  2.0 & 2.8  & 2.0 & 2.8 \\
      Si & Cu &  0.14 & 2.46 & 4.74 & 0.0 & 2.8  &  0.0 & 2.8  &  2.0 & 2.8  & 2.0 & 2.8 \\
      Si & Fe & -0.07 & 2.39 & 5.17 & 1.0 & 2.8  &  1.0 & 2.8  &  2.0 & 2.8  & 0.0 & 2.8 \\
      Mg & Cu &  0.23 & 2.63 & 4.70 & 2.0 & 2.8  &  0.0 & 2.8  &  2.0 & 2.8  & 2.0 & 2.8 \\
      Mg & Fe &  0.60 & 2.61 & 4.96 & 0.65& 2.8  &  0.0 & 2.8  &  2.0 & 2.8  & 2.0 & 2.8 \\
      Cu & Fe &  0.63 & 2.42 & 5.21 & 2.0 & 2.8  &  0.0 & 2.8  &  2.0 & 2.8  & 2.0 & 2.8 \\
    \end{tabular*}
  \end{ruledtabular}
\end{table*}

\section{MEAM parameters for element pairs}
\label{sec:pairs}

The MEAM potential parameters for each element pair were
initialized to match the ab-initio heat of formation, equilibrium
volume, bulk modulus, and elastic moduli in the hypothetical NaCl
reference structure, which was chosen for its simplicity. 
Since the equilibrium volume, cohesive energy,
and bulk modulus of the NaCl structure are directly related to MEAM
parameters, they can be reproduced exactly. An improved agreement of
the shear moduli from MEAM and ab-initio simulations was achieved in
some cases by adjusting the electron density scaling factor $\rho_0$.
Then, heat of formation of binary compounds in a variety of crystal
structures from MEAM were examined and compared with the ab-initio
results. To correlate the MEAM results with the lowest formation energies
of the compounds from DFT calculations, the MEAM screening and
$\triangle H_{\text{B1}}^{\text{XY}}$ parameters for element pairs
were adjusted. The final MEAM parameters are given in
Table~\ref{tab:meam_parp}. The predicted MEAM properties for the NaCl
reference structure are compared with DFT results in
Table~\ref{tab:nacl}, and show that in general the MEAM heat of
formation, bulk modulus, and equilibrium volume reproduce the DFT
results well.  In contrast, the shear elastic constants are not well
reproduced.  In fact the sign of the shear elastic constant,
representing crystal stability, is frequently in disagreement with
the DFT results.  This is really not a significant problem as the NaCl structure
does not exist in nature.  A more important criteria for success of
these potentials is how they perform for lower energy crystal
structures.  We address this issue in the next section.

\begingroup
\squeezetable
\begin{table}[]
  \caption{\label{tab:nacl}Structural and elastic properties of
    element pairs in the reference NaCl (B1) crystal structure
    from DFT and MEAM calculations.
    $\Delta H$ is the heat of formation in eV/atom,
    $V_0$ is the volume per atom in \AA$^3$.
    Elastic constants $B_0$, $C_{44}$, and $(C_{11}-C_{12})/2$
    are in GPa.
  }
  \begin{ruledtabular}
    \begin{tabular}{ccccccc}
      pair & method & $\Delta H$ & $V_0$ & $B_0$ & $C_{44}$ & $\frac{C_{11}-C_{12}}{2}$\\
      \hline
      \multirow{2}{*}{AlSi}
      & DFT  &  0.28 & 17.9 &   76.7 &   10 &   76 \\
      & MEAM &  0.28 & 18.0 &   76.4 &  -13 &    8 \\
      \hline
      \multirow{2}{*}{AlMg}
      & DFT  &  0.42 & 23.7 &   30.9 &  -18 &   36 \\
      & MEAM &  0.23 & 23.6 &   33.9 &   -3 &   35 \\
      \hline
      \multirow{2}{*}{AlCu}
      & DFT  &  0.19 & 16.1 &   77.5 &  -18 &   52 \\
      & MEAM &  0.19 & 16.2 &   77.4 &  -19 &   56 \\
      \hline
      \multirow{2}{*}{AlFe}
      & DFT  &  0.36 & 14.7 &   90.3 &  -25 &  105 \\
      & MEAM &  0.26 & 14.7 &   92.7 &  -27 &  109 \\
      \hline
      \multirow{2}{*}{SiMg}
      & DFT  &  0.41 & 20.9 &   50.6 &  -26 &   48 \\
      & MEAM &  0.20 & 20.8 &   54.9 &    9 &   61 \\
      \hline
      \multirow{2}{*}{SiCu}
      & DFT  &  0.39 & 14.9 &   99.0 &  -29 &   58 \\
      & MEAM &  0.14 & 14.9 &  105.9 &    9 &  223 \\
      \hline
      \multirow{2}{*}{SiFe}
      & DFT  &  0.25 & 12.9 &  100.9 &  -70 &  112 \\
      & MEAM & -0.07 & 13.7 &  157.9 &   65 &  363 \\
      \hline
      \multirow{2}{*}{MgCu}
      & DFT  &  0.23 & 18.5 &   48.7 &  -10 &   49 \\
      & MEAM &  0.23 & 18.2 &   49.6 &   -1 &   61 \\
      \hline
      \multirow{2}{*}{MgFe}
      & DFT  &  0.86 & 17.7 &   50.4 &  -23 &   83 \\
      & MEAM &  0.60 & 17.8 &   56.5 &  -17 &   62 \\
      \hline
      \multirow{2}{*}{CuFe}
      & DFT  &  0.78 & 14.1 &  107.4 &  -23 &  134 \\
      & MEAM &  0.63 & 14.2 &  111.8 &   10 &  131 \\
    \end{tabular}
  \end{ruledtabular}
\end{table}
\endgroup

\subsection{Heat of formation for binary compounds}

The alloy phases that the MEAM potential predicts as most likely to
form at the temperature $T=0$~K are those with the lowest heat of
formation per atom, $\Delta H$, which is defined as
\begin{equation}
  \Delta H =
  \frac
  {E_{\text{tot}}[N_{\text{X}} + N_{\text{Y}}]
    - N_{\text{X}}\varepsilon_{\text{X}} 
    - N_{\text{Y}}\varepsilon_{\text{Y}}
  }
  {N_{\text{X}} + N_{\text{Y}}
  },
  \label{eq:hof}
\end{equation}
$E_{\text{tot}}$ is the total energy of the simulation cell,
$N_{\text{X}}$ and $N_{\text{Y}}$ are the numbers of type-X
and type-Y atoms in the cell, $\varepsilon_{\text{X}}$ and
$\varepsilon_{\text{Y}}$ are the total energies per atom for
type-X and type-Y in their ground state bulk structures,
respectively.

\begin{figure*}[!htbp]
  \includegraphics[]{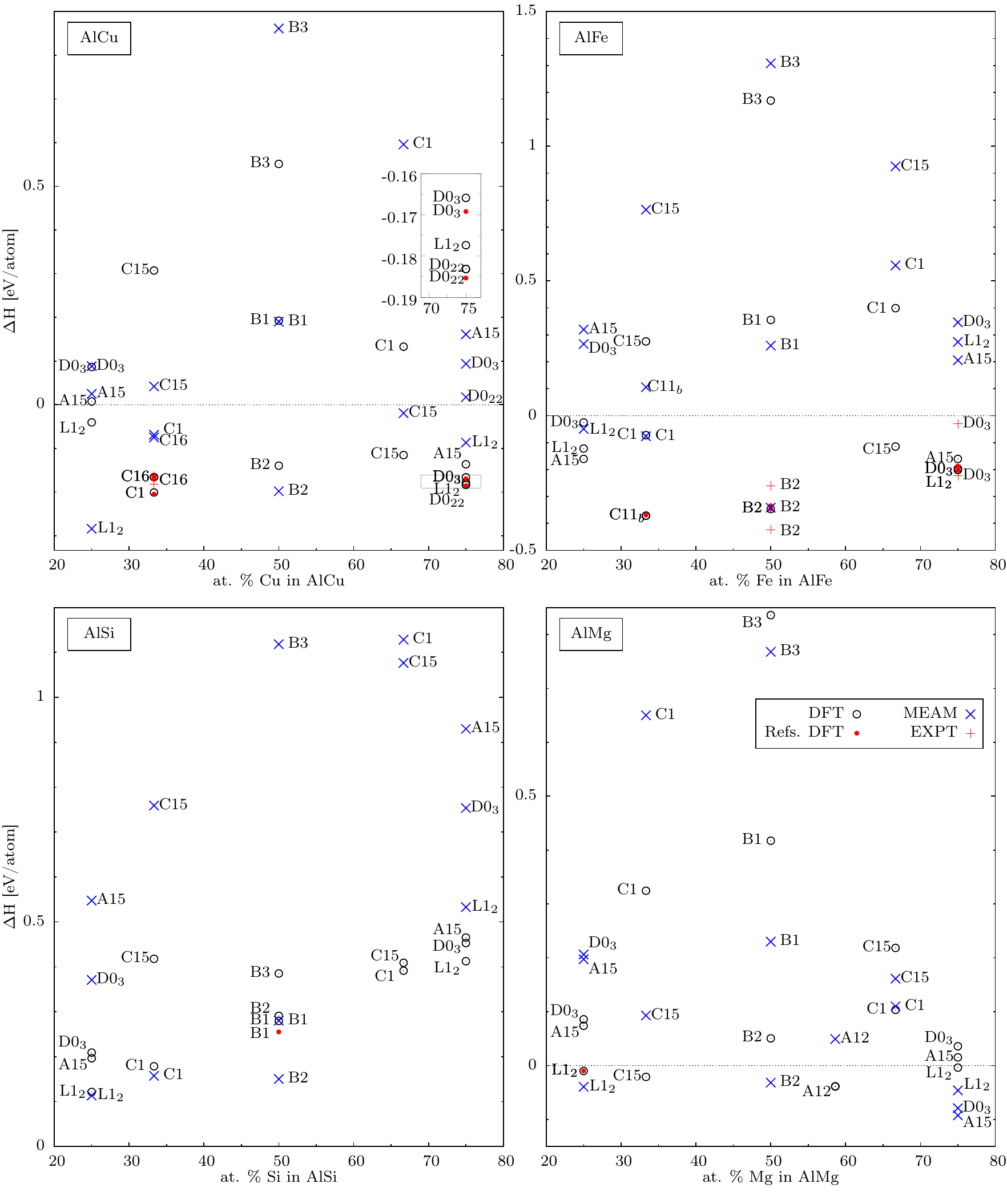}
  \caption{(Color online) Heat of formation of AlSi, AlMg, AlCu, AlFe
    binary compounds from MEAM, DFT, and experiments.
    References:
    AlSi~\cite{gall00:atomis},
    AlMg~\cite{site:cmu},
    AlCu~\cite{ravi06:pred,wolv01:entro_al2cu,murray85:alcu,ravi05:compar_al_alloys},
    AlFe~\cite{besson97:dev_pot_feal,shaojun98:fpvib_feal,zhang97:inter_inv}.
    DFT points are labeled on the left, MEAM and experimental on the right.
    Values for the most stable compounds are also shown in
    Table~\ref{tab:elast1}.
    The inside plot is a magnified portion of a larger plot.
  }
  \label{fig:hofs4}
\end{figure*}

\begin{figure*}[!htbp]
  \includegraphics[]{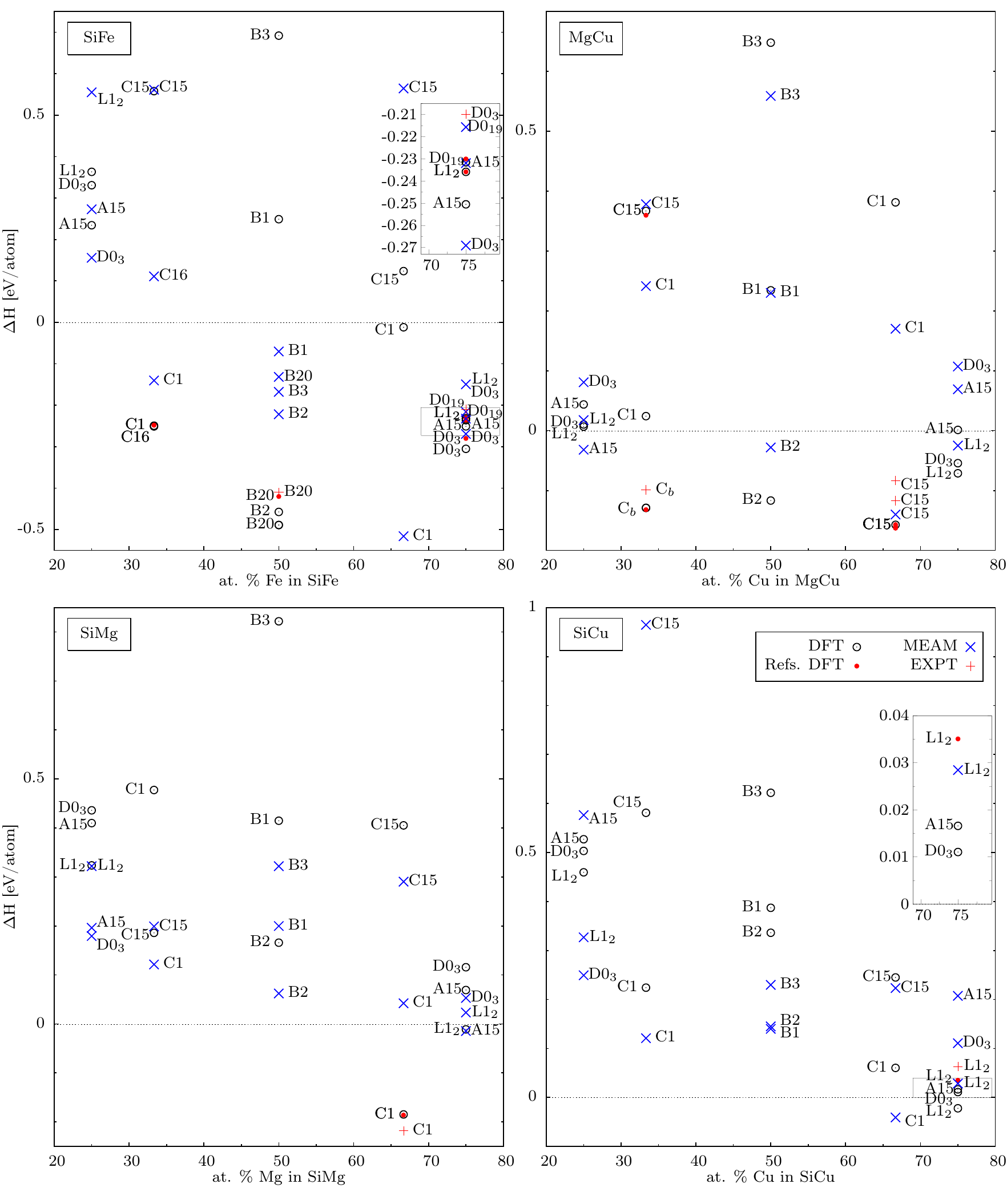}
  \caption{(Color online) Heat of formation of SiMg, SiCu, SiFe, MgCu
    binary compounds from MEAM, DFT, and experiments.
    References:
    SiMg~\cite{ravi05:compar_al_alloys},
    SiCu~\cite{meschel91:note_eof_cu3si,solares06:multi},
    SiFe~\cite{moroni99:coh,site:cmu}, and
    MgCu~\cite{zhou07:mod_cumg,site:cmu}.
    DFT points are labeled on the left, MEAM and experimental on the right.
    Values for the most stable compounds are also shown in
    Table~\ref{tab:elast2}.
    The inside plot is a magnified portion of a larger plot.
  }
  \label{fig:hofs5}
\end{figure*}

\begin{figure*}[!htbp]
  \includegraphics[]{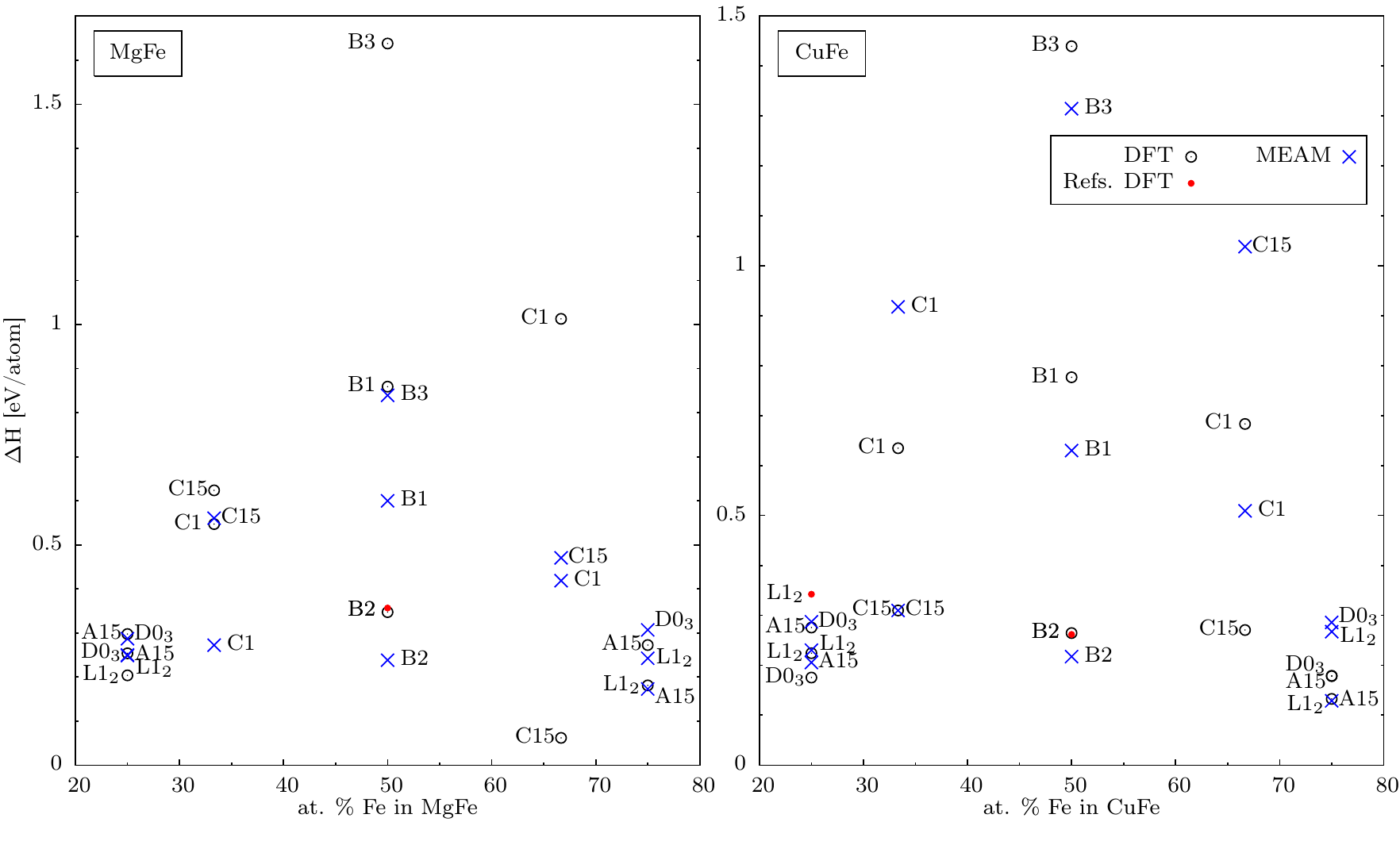}
  \caption{(Color online) Heat of formation of MgFe and CuFe
    binary compounds from MEAM and DFT.
    References:
    MgFe~\cite{site:cmu},
    CuFe~\cite{site:cmu}.
    DFT points are labeled on the left, MEAM on the right.
    Values for the most stable compounds are also shown in
    Table~\ref{tab:elast2}.
  }
  \label{fig:hofs6}
\end{figure*}

To check the validity of our new potentials, we computed the heat of
formation per atom for many intermetallic phases of all alloy pairs.
The total energy values in Eq.~(\ref{eq:hof}) for B1, B2, B3, C1, C15,
D0$_3$, A15, L1$_2$, and other relevant structures were evaluated at
the optimal atomic volume for each structure. Heat of formation for
basic binary compounds based on the new MEAM potential and DFT results were
calculated and compared with experimental values
(Figures~\ref{fig:hofs4}--\ref{fig:hofs6}). The DFT and MEAM results
for the phases with lowest $\Delta H$ are also shown in
Tables~\ref{tab:elast1}--\ref{tab:elast2}.

The agreement between MEAM and DFT is quite satisfactory. In most
cases, the MEAM results preserve the order of stability predicted by
the DFT results. The
differences in the heat of formation per atom from the MEAM and DFT results are
less than 0.5~eV at most. In general the atomic volumes predicted by
MEAM agree at least qualitatively with the DFT and experimental results.
The MEAM calculations of the bulk moduli also agree semi-quantitatively with
DFT and experimental results, usually within 20\%. Predicted shear moduli 
usually follow the DFT and experimental results, but in some cases there
is significant disagreement.

\begingroup
\squeezetable
\begin{table}[!t]
\caption{The formation energies of substitutional point defects in Al, Si, Mg, Cu, and Fe. All energy values are given in eV. DFT values are given in parentheses.\label{tab:subst}}
\begin{ruledtabular}
\begin{tabular}{cccccc}
\multirow{2}{*}{Host} & \multicolumn{5}{c}{Substitute atom} \\
\cline{2-6}
    & Al           & Si             & Mg          & Cu               & Fe         \\
\hline
Al  &              &  0.5 (0.5)     &$-$0.2 (0.05)& $-$1.1 ($-$0.1)  &$-$1.3 ($-$0.4)\\
Si  &  7.0  (0.9)  &                & 2.9 (2.4)   &  1.9   (2.)     & 1.6  (1.9)   \\
Mg  &$-$0.7 (0.06) & 0.2 (0.4)      &             & $-$0.2 (0.2)     & 1.5  (1.1)   \\
Cu  & 0.7 ($-$0.7) & 0.8 ($-$0.2)   &1.1 ($-$0.2) &                  & 2.9  (1.4)  \\
Fe  & 0.2 ($-$0.7) &$-$2.9 ($-$1.1) &0.8 (1.0)    & $-$0.3 (0.8)     &              \\   
\end{tabular}
\end{ruledtabular}
\end{table}
\endgroup

\setlength{\extrarowheight}{2pt}
\begingroup
\squeezetable
\begin{table*}[!tbhp]
   \caption{\label{tab:elast1}Structural and elastic properties of
     element pairs in varying crystal structures
     from the present DFT and MEAM calculations compared with references and measured values.
     $\Delta H$ is the heat of formation in meV/atom, $V_0$ is the volume per atom in \AA$^3$, and elastic constants $B_0$, $C_{44}$, and $(C_{11}-C_{12})/2$ are in GPa.
   }
  \begin{ruledtabular}
    \begin{tabular}{lllHVCCC}
      compos. & str. & met. & $\Delta H$ & $V_0$ & $B_0$ & $C_{44}$ &$\frac{C_{11}-C_{12}}{2}$\\\hline
      Al$_{3}$Si$_{}$ & L1$_2$ & DFT & 121 & 16.04 & 74.3 & 24.1 & 9.4 \tabularnewline
         &          & MEAM/EAM & 113 & 16.67 & 96.7 & 31.2 & 31.2 \tabularnewline
\cline{2-8}

Al$_{2}$Si$_{}$ & C1 & DFT & 178 & 18.78 & 62.9 & 25.4 & $-$11.8 \tabularnewline
         &          & MEAM/EAM & 157 & 19.17 & 73.6 & 15.3 & 0.0 \tabularnewline
\cline{2-8}

Al$_{}$Si$_{}$ & B2 & DFT & 291 & 15.91 & 78.8 & 22.4 & $-$32.8 \tabularnewline
         &          & MEAM/EAM & 150 & 16.25 & 102.1 & 29.1 & $-$17.0 \tabularnewline
\cline{2-8}

\hline

      Al$_{12}$Mg$_{17}$ & A12 & DFT & $-$39, $-$35\cite{narasimhan95:ab-poly-almg}, $-$37\cite{zhong05:contr_fp_almg}, $-$48\cite{wang08:str_mg17al12_fp} & 20.12, 18.65\cite{narasimhan95:ab-poly-almg}, 20.04\cite{zhong05:contr_fp_almg}, 20.25\cite{wang08:str_mg17al12_fp} & 50.1, 49.6\cite{wang08:str_mg17al12_fp} & 20.0\cite{wang08:str_mg17al12_fp} & 28.9\cite{wang08:str_mg17al12_fp} \tabularnewline
         &          & MEAM/EAM & 49, $-$19\cite{kim09:atom_mgal}, $-$36\cite{liu97:anis_almg} & 21.28, 20.30\cite{kim09:atom_mgal} & 49.6\cite{kim09:atom_mgal} & 11.5\cite{kim09:atom_mgal} & 23.9\cite{kim09:atom_mgal} \tabularnewline
         &          & EXP & $-$34\cite{murray1982:almg} & 20.13\cite{villars85:pearson}, 20.30\cite{singh03:ext_mgal} &  &  &  \tabularnewline
\cline{2-8}

Al$_{3}$Mg$_{}$ & L1$_2$ & DFT & $-$10, $-$15\cite{narasimhan95:ab-poly-almg}, $-$9\cite{mendelev09:dev_pot_s-l_almg} & 17.80, 16.52\cite{narasimhan95:ab-poly-almg}, 17.78\cite{mendelev09:dev_pot_s-l_almg} & 63.3 & 35.3 & 31.9 \tabularnewline
         &          & MEAM/EAM & $-$39, $-$2\cite{mendelev09:dev_pot_s-l_almg} & 19.04, 17.54\cite{mendelev09:dev_pot_s-l_almg} & 62.7 & 33.3 & 14.7 \tabularnewline
\cline{2-8}

Al$_{}$Mg$_{3}$ & L1$_2$ & DFT & $-$4, $-$5\cite{narasimhan95:ab-poly-almg}, $-$3\cite{mendelev09:dev_pot_s-l_almg} & 20.98, 19.18\cite{narasimhan95:ab-poly-almg}, 19.89\cite{mendelev09:dev_pot_s-l_almg} & 43.9 & 25.2 & 22.6 \tabularnewline
         &          & MEAM/EAM & $-$46, 21\cite{mendelev09:dev_pot_s-l_almg} & 21.99, 20.98\cite{mendelev09:dev_pot_s-l_almg} & 45.0 & 24.9 & 13.8 \tabularnewline
\cline{2-8}

Al$_{}$Mg$_{}$ & B2 & DFT & 51, 50\cite{narasimhan95:ab-poly-almg}, 51\cite{narasimhan95:ab-poly-almg} & 19.50, 18.00\cite{narasimhan95:ab-poly-almg}, 19.48\cite{narasimhan95:ab-poly-almg} & 47.9 & 38.4 & $-$7.5 \tabularnewline
         &          & MEAM/EAM & $-$32, 90\cite{narasimhan95:ab-poly-almg} & 20.29, 19.57\cite{narasimhan95:ab-poly-almg} & 50.6 & 32.9 & 8.9 \tabularnewline
\cline{2-8}

\hline

      Al$_{2}$Cu$_{}$ & C16 & DFT & $-$166, $-$170\cite{site:cmu}, $-$166\cite{ravi05:compar_al_alloys}, $-$163\cite{ravi06:pred} & 14.95, 14.89\cite{ravi06:pred} & 96.3 &  &  \tabularnewline
         &          & MEAM/EAM & $-$75 & 15.80 &  &  &  \tabularnewline
         &          & EXP & $-$161\cite{fries98:compil_cost} & 14.90\cite{ravi06:pred} &  &  &  \tabularnewline
\cline{2-8}

         & C1 & DFT & $-$201, $-$202\cite{ravi06:pred}, $-$204\cite{site:cmu} & 16.14, 16.12\cite{ravi06:pred} & 93.1 & 81.6 & 46.0 \tabularnewline
         &          & MEAM/EAM & $-$69 & 17.31 & 77.4 & 43.2 & 7.7 \tabularnewline
         &          & EXP &  & 15.63\cite{ravi06:pred} &  &  &  \tabularnewline
\cline{2-8}

Al$_{}$Cu$_{3}$ & D0$_{22}$ & DFT & $-$183, $-$185\cite{site:cmu} & 12.52 & 128.1 &  &  \tabularnewline
         &          & MEAM/EAM & 17 & 13.08 &  &  &  \tabularnewline
\cline{2-8}

         & L1$_2$ & DFT & $-$177, $-$219\cite{liu:diss} & 12.59 & 128.4 & 81.1 & 13.8 \tabularnewline
         &          & MEAM/EAM & $-$87, $-$229\cite{liu:diss} & 13.47 & 131.9 & 72.4 & 26.8 \tabularnewline
\cline{2-8}

         & D0$_3$ & DFT & $-$166, $-$169\cite{site:cmu} & 12.60 & 127.9 & 98.4 & 1.1 \tabularnewline
         &          & MEAM/EAM & 94 & 12.34 & 107.8 & 88.7 & 4.9 \tabularnewline
\cline{2-8}

         & A15 & DFT & $-$136 & 12.77 & 124.7 & 34.5 & 102.8 \tabularnewline
         &          & MEAM/EAM & 161 & 13.48 & 111.6 & 200.5 & 92.8 \tabularnewline
\cline{2-8}

Al$_{3}$Cu$_{}$ & L1$_2$ & DFT & $-$40 & 15.22 & 89.9 & 23.4 & 61.5 \tabularnewline
         &          & MEAM/EAM & $-$284 & 14.58 & 106.0 & 40.3 & 41.2 \tabularnewline
\cline{2-8}

Al$_{}$Cu$_{}$ & B2 & DFT & $-$139 & 13.45 & 108.6 & 31.4 & $-$15.4 \tabularnewline
         &          & MEAM/EAM & $-$198 & 14.15 & 109.2 & 56.7 & 2.8 \tabularnewline
\cline{2-8}

\hline

      Al$_{}$Fe$_{}$ & B2 & DFT & $-$347, $-$347\cite{site:cmu}, $-$334\cite{maugis2006:ab_fealc}, $-$379\cite{gonzales02:ab_form_bcc}, $-$420\cite{watson98:tm_al}, $-$400\cite{kulikov99:onset_mag_b2}, $-$338\cite{friak10:ab_vol_feal}, $-$311\cite{lechermann05:fp_nialfe}, $-$1500\cite{shaojun98:fpvib_feal} & 11.86, 11.88\cite{maugis2006:ab_fealc}, 12.07\cite{gonzales02:ab_form_bcc}, 11.33\cite{fu92:def_b2_alum}, 11.22\cite{watson98:tm_al}, 11.65\cite{kulikov99:onset_mag_b2}, 11.89\cite{friak10:ab_vol_feal}, 11.93\cite{lechermann05:fp_nialfe}, 12.19\cite{shaojun98:fpvib_feal} & 174.6, 177.0\cite{maugis2006:ab_fealc}, 183.0\cite{fu92:def_b2_alum}, 155.0\cite{lechermann05:fp_nialfe}, 156.0\cite{shaojun98:fpvib_feal} & 138.8, 165.0\cite{fu92:def_b2_alum}, 107.0\cite{shaojun98:fpvib_feal} & 61.8, 80.0\cite{fu92:def_b2_alum}, 38.1\cite{shaojun98:fpvib_feal} \tabularnewline
         &          & MEAM/EAM & $-$342, $-$298\cite{lee10:meam_fe-al}, $-$260\cite{besson97:dev_pot_feal}, $-$106\cite{shu01:vac} & 12.88, 12.32\cite{vailhe97:shear_b2_feal}, 13.92\cite{shu01:vac}, 11.45\cite{zhang97:inter_inv} & 145.3, 144.0\cite{vailhe97:shear_b2_feal}, 138.0\cite{shu01:vac}, 193.0\cite{zhang97:inter_inv} & 111.7, 117.0\cite{vailhe97:shear_b2_feal}, 110.7\cite{shu01:vac} & 79.6, 18.0\cite{vailhe97:shear_b2_feal}, 7.9\cite{shu01:vac} \tabularnewline
         &          & EXP & $-$260\cite{kubaschewski55:hofs_tial_tife}, $-$423\cite{desai87:therm_bin_al}, $-$250\cite{hultgren73:selected}, $-$280\cite{gale04:smithells} & 12.23\cite{pearson58:handbook} & 152.0\cite{yoo90:slip_b2}, 136.0\cite{simmons71:single} & 127.0\cite{yoo90:slip_b2}, 127.1\cite{simmons71:single} & 43.0\cite{yoo90:slip_b2}, 33.7\cite{simmons71:single} \tabularnewline
\cline{2-8}

Al$_{}$Fe$_{3}$ & D0$_3$ & DFT & $-$202, $-$203\cite{site:cmu}, $-$200\cite{maugis2006:ab_fealc}, $-$221\cite{gonzales02:ab_form_bcc}, $-$230\cite{watson98:tm_al}, $-$202\cite{friak10:ab_vol_feal}, $-$201\cite{lechermann05:fp_nialfe}, $-$1860\cite{shaojun98:fpvib_feal} & 11.79, 11.81\cite{maugis2006:ab_fealc}, 12.01\cite{gonzales02:ab_form_bcc}, 14.65\cite{watson98:tm_al}, 11.82\cite{friak10:ab_vol_feal}, 12.09\cite{lechermann05:fp_nialfe}, 11.57\cite{shaojun98:fpvib_feal} & 160.0, 174.0\cite{maugis2006:ab_fealc}, 151.0\cite{lechermann05:fp_nialfe}, 170.0\cite{shaojun98:fpvib_feal} & 140.0, 137.5\cite{shaojun98:fpvib_feal} & 25.5, 10.8\cite{shaojun98:fpvib_feal} \tabularnewline
         &          & MEAM/EAM & 346, $-$206\cite{lee10:meam_fe-al}, $-$222\cite{besson97:dev_pot_feal}, $-$74\cite{shu01:vac} & 12.01, 11.77\cite{besson97:dev_pot_feal}, 12.86\cite{shu01:vac}, 10.80\cite{zhang97:inter_inv} & 137.5, 146.0\cite{shu01:vac}, 229.0\cite{zhang97:inter_inv} & 129.0, 162.0\cite{besson97:dev_pot_feal}, 126.3\cite{shu01:vac} & 30.0, 53.0\cite{besson97:dev_pot_feal}, 12.6\cite{shu01:vac} \tabularnewline
         &          & EXP & $-$202\cite{hultgren73:selected}, $-$321\cite{desai87:therm_bin_al} & 12.07\cite{pearson58:handbook} & 144.0\cite{simmons71:single} & 131.7\cite{simmons71:single} & 20.2\cite{simmons71:single} \tabularnewline
\cline{2-8}

         & L1$_2$ & DFT & $-$196, $-$187\cite{site:cmu}, $-$200\cite{maugis2006:ab_fealc}, $-$40\cite{watson98:tm_al}, $-$222\cite{lechermann05:fp_nialfe} & 12.13, 12.14\cite{maugis2006:ab_fealc}, 14.14\cite{watson98:tm_al}, 12.35\cite{lechermann05:fp_nialfe} & 166.3, 158.0\cite{maugis2006:ab_fealc}, 168.0\cite{lechermann05:fp_nialfe} & 125.1 & 11.5 \tabularnewline
         &          & MEAM/EAM & 274, $-$180\cite{lee10:meam_fe-al} & 12.88 & 139.5 & 96.9 & 84.0 \tabularnewline
\cline{2-8}

         & A15 & DFT & $-$161 & 12.08 & 156.9 & 67.3 & 135.1 \tabularnewline
         &          & MEAM/EAM & 205 & 12.59 & 166.7 & 35.5 & 153.5 \tabularnewline
\cline{2-8}

Al$_{}$Fe$_{2}$ & C15 & DFT & $-$115, $-$60\cite{watson98:tm_al} & 12.42, 11.43\cite{watson98:tm_al} & 130.2 & 55.0 & 52.2 \tabularnewline
         &          & MEAM/EAM & 925 & 13.67 & 127.8 & 259.2 & 551.0 \tabularnewline
\cline{2-8}

Al$_{3}$Fe$_{}$ & A15 & DFT & $-$161 & 13.91 & 121.5 & 67.7 & 120.0 \tabularnewline
         &          & MEAM/EAM & 321 & 15.03 & 103.5 & 1.8 & 66.6 \tabularnewline
\cline{2-8}

         & L1$_2$ & DFT & $-$122, $-$150\cite{watson98:tm_al}, $-$105\cite{lechermann05:fp_nialfe} & 13.68, 13.07\cite{watson98:tm_al}, 13.69\cite{lechermann05:fp_nialfe} & 126.5, 98.8\cite{lechermann05:fp_nialfe} & 85.9 & 49.9 \tabularnewline
         &          & MEAM/EAM & $-$49 & 14.83 & 108.5 & 59.4 & 20.3 \tabularnewline
\cline{2-8}

         & D0$_3$ & DFT & $-$25, $-$99\cite{gonzales02:ab_form_bcc}, $-$13\cite{lechermann05:fp_nialfe} & 13.38, 13.57\cite{gonzales02:ab_form_bcc}, 13.35\cite{lechermann05:fp_nialfe} & 126.0, 119.6\cite{lechermann05:fp_nialfe} & 91.4 & $-$48.3 \tabularnewline
         &          & MEAM/EAM & 266 & 14.81 & 93.8 & 57.7 & $-$31.1 \tabularnewline
\cline{2-8}

Al$_{2}$Fe$_{}$ & C11$_b$ & DFT & $-$371, $-$420\cite{watson98:tm_al} & 12.78, 12.35\cite{watson98:tm_al}, 12.80\cite{krajci02:cov_bond_trans_al} & 149.0 &  &  \tabularnewline
         &          & MEAM/EAM & 106 & 14.71 &  &  &  \tabularnewline
\cline{2-8}

         & C1 & DFT & $-$72 & 15.25 & 98.6 & 76.8 & 55.0 \tabularnewline
         &          & MEAM/EAM & $-$76 & 16.12 & 90.4 & 47.5 & 36.4 \tabularnewline
\cline{2-8}

\hline

    \end{tabular}
  \end{ruledtabular}
\end{table*}
\endgroup

\begingroup
\squeezetable
\begin{table*}[!tbhp]
  \caption{\label{tab:elast2}Structural and elastic properties of
    element pairs in varying crystal structures
    from the present DFT and MEAM calculations compared with references and measured values.
    $\Delta H$ is the heat of formation in meV/atom, $V_0$ is the volume per atom in \AA$^3$, and elastic constants $B_0$, $C_{44}$, and $(C_{11}-C_{12})/2$ are in GPa.
  }
  \begin{ruledtabular}
    \begin{tabular}{lllHVCCC}
      compos. & str. & met. & $\Delta H$ & $V_0$ & $B_0$ & $C_{44}$ &$\frac{C_{11}-C_{12}}{2}$\\\toprule
      Si$_{}$Mg$_{2}$ & C1 & DFT & $-$185, $-$186\cite{ravi05:compar_al_alloys} & 21.41 & 54.1 & 47.6 & 47.3 \tabularnewline
         &          & MEAM/EAM & 42 & 23.05 & 47.8 & 20.9 & 32.2 \tabularnewline
         &          & EXP & $-$225\cite{fries98:compil_cost} &  &  &  &  \tabularnewline
\cline{2-8}

Si$_{}$Mg$_{3}$ & L1$_2$ & DFT & $-$11 & 19.29 & 50.8 & 29.9 & 37.1 \tabularnewline
         &          & MEAM/EAM & 24 & 20.70 & 57.5 & 23.5 & 21.8 \tabularnewline
\cline{2-8}

         & A15 & DFT & 69 & 20.09 & 44.1 & 9.3 & 31.8 \tabularnewline
         &          & MEAM/EAM & $-$14 & 21.25 & 56.3 & 32.2 & 33.9 \tabularnewline
\cline{2-8}

\hline

      Si$_{}$Cu$_{3}$ & L1$_2$ & DFT & $-$22, 35\cite{solares06:multi} & 12.18 & 137.3 & 65.0 & 38.1 \tabularnewline
         &          & MEAM/EAM & 28 & 13.37 & 134.9 & 64.4 & 33.0 \tabularnewline
         &          & EXP & 63\cite{meschel91:note_eof_cu3si} &  &  &  &  \tabularnewline
\cline{2-8}

Si$_{}$Cu$_{2}$ & C1 & DFT & 60 & 14.26 & 111.9 & 76.3 & 23.1 \tabularnewline
         &          & MEAM/EAM & $-$41 & 14.81 & 102.8 & 97.9 & 16.2 \tabularnewline
\cline{2-8}

\hline

      Si$_{}$Fe$_{}$ & B20 & DFT & $-$489, $-$484\cite{site:cmu}, $-$420\cite{moroni99:coh} & 11.04 & 226.5 &  &  \tabularnewline
         &          & MEAM/EAM & $-$132 & 13.11 &  &  &  \tabularnewline
         &          & EXP & $-$410\cite{moroni99:coh} &  &  &  &  \tabularnewline
\cline{2-8}

         & B2 & DFT & $-$457 & 10.55 & 231.9 & 87.0 & 155.8 \tabularnewline
         &          & MEAM/EAM & $-$222 & 13.09 & 177.7 & 36.2 & 225.3 \tabularnewline
\cline{2-8}

Si$_{}$Fe$_{3}$ & D0$_3$ & DFT & $-$305, $-$315\cite{site:cmu}, $-$280\cite{moroni99:coh} & 10.99 & 204.5 & 142.4 & 54.5 \tabularnewline
         &          & MEAM/EAM & $-$269 & 12.03 & 169.2 & 91.6 & 36.6 \tabularnewline
         &          & EXP & $-$210\cite{moroni99:coh} &  &  &  &  \tabularnewline
\cline{2-8}

         & A15 & DFT & $-$251 & 11.44 & 173.3 & 72.3 & 125.3 \tabularnewline
         &          & MEAM/EAM & $-$232 & 12.28 & 190.1 & 47.6 & 119.9 \tabularnewline
\cline{2-8}

         & L1$_2$ & DFT & $-$236, $-$236\cite{site:cmu} & 11.38 & 188.4 & 116.0 & 26.6 \tabularnewline
         &          & MEAM/EAM & $-$149 & 11.80 & 188.6 & 65.1 & 135.8 \tabularnewline
\cline{2-8}

         & D0$_{19}$ & DFT & $-$232, $-$230\cite{site:cmu} & 11.28 & 160.1 &  &  \tabularnewline
         &          & MEAM/EAM & $-$216 & 12.05 &  &  &  \tabularnewline
\cline{2-8}

Si$_{2}$Fe$_{}$ & C1 & DFT & $-$249, $-$248\cite{site:cmu} & 13.05 & 169.4 & 136.8 & 15.5 \tabularnewline
         &          & MEAM/EAM & $-$140 & 15.53 & 158.7 & 95.7 & 41.3 \tabularnewline
\cline{2-8}

         & C16 & DFT & $-$251, $-$248\cite{site:cmu} & 12.39 & 170.6 &  &  \tabularnewline
         &          & MEAM/EAM & 111 & 13.64 &  &  &  \tabularnewline
\cline{2-8}

Si$_{}$Fe$_{2}$ & C1 & DFT & $-$12 & 12.95 & 159.6 & 82.6 & $-$1.4 \tabularnewline
         &          & MEAM/EAM & $-$516 & 12.95 & 182.4 & 154.6 & 226.8 \tabularnewline
\cline{2-8}

\hline

      Mg$_{2}$Cu$_{}$ & C$_b$ & DFT & $-$129, $-$131\cite{site:cmu}, $-$132\cite{bailey04:sim_cumg}, $-$137\cite{zhou07:mod_cumg} & 18.13 & 57.1 &  &  \tabularnewline
         &          & EXP & $-$99\cite{king64:a_thermoch_laves} &  &  &  &  \tabularnewline
\cline{2-8}

Mg$_{}$Cu$_{2}$ & C15 & DFT & $-$157, $-$160\cite{site:cmu}, $-$163\cite{zhou07:mod_cumg}, $-$157\cite{bailey04:sim_cumg} & 14.58 & 90.6 & 45.4 & 25.2 \tabularnewline
         &          & MEAM/EAM & $-$140 & 14.50 & 104.7 & 21.5 & $-$12.0 \tabularnewline
         &          & EXP & $-$117\cite{king64:a_thermoch_laves}, $-$83\cite{predel72:beitrag} &  &  &  &  \tabularnewline
\cline{2-8}

Mg$_{}$Cu$_{}$ & B2 & DFT & $-$117 & 15.65 & 69.3 & 60.3 & 16.9 \tabularnewline
         &          & MEAM/EAM & $-$28 & 15.56 & 73.6 & 51.1 & $-$11.6 \tabularnewline
\cline{2-8}

Mg$_{}$Cu$_{3}$ & L1$_2$ & DFT & $-$71 & 13.49 & 96.6 & 62.6 & 21.5 \tabularnewline
         &          & MEAM/EAM & $-$25 & 13.81 & 103.9 & 42.5 & 29.0 \tabularnewline
\cline{2-8}

         & D0$_3$ & DFT & $-$54 & 13.52 & 96.5 & 75.2 & 0.3 \tabularnewline
         &          & MEAM/EAM & 107 & 13.43 & 95.7 & 64.1 & $-$22.4 \tabularnewline
\cline{2-8}

\hline

      Mg$_{}$Fe$_{2}$ & C15 & DFT & 62 & 13.58 & 91.4 & 72.0 & 53.4 \tabularnewline
         &          & MEAM/EAM & 471 & 14.79 & 95.8 & 322.9 & 605.7 \tabularnewline
\cline{2-8}

Mg$_{}$Fe$_{3}$ & L1$_2$ & DFT & 181 & 13.11 & 122.6 & 96.9 & 14.0 \tabularnewline
         &          & MEAM/EAM & 243 & 13.65 & 116.4 & 64.5 & 45.5 \tabularnewline
\cline{2-8}

Mg$_{3}$Fe$_{}$ & L1$_2$ & DFT & 204 & 18.29 & 51.6 & 52.1 & 18.9 \tabularnewline
         &          & MEAM/EAM & 249 & 19.08 & 54.5 & 30.0 & 23.1 \tabularnewline
\cline{2-8}

Mg$_{}$Fe$_{}$ & B2 & DFT & 347, 357\cite{site:cmu} & 15.39 & 71.4 & 68.7 & $-$25.4 \tabularnewline
         &          & MEAM/EAM & 238 & 15.82 & 86.4 & 63.2 & 13.7 \tabularnewline
\cline{2-8}

\hline

      Cu$_{}$Fe$_{3}$ & L1$_2$ & DFT & 133 & 11.73 & 132.5 & 99.9 & 6.8 \tabularnewline
         &          & MEAM/EAM & 267 & 11.78 & 172.6 & 80.5 & 68.0 \tabularnewline
\cline{2-8}

         & A15 & DFT & 178 & 11.95 & 137.6 & 54.2 & 134.7 \tabularnewline
         &          & MEAM/EAM & 129 & 12.26 & 192.3 & 70.5 & 155.2 \tabularnewline
\cline{2-8}

Cu$_{3}$Fe$_{}$ & D0$_3$ & DFT & 175 & 12.10 & 139.2 & 105.5 & $-$2.0 \tabularnewline
         &          & MEAM/EAM & 287 & 11.59 & 134.0 & 130.8 & 20.7 \tabularnewline
\cline{2-8}

         & L1$_2$ & DFT & 224, 342\cite{site:cmu} & 12.16 & 135.2 & 60.7 & $-$0.0 \tabularnewline
         &          & MEAM/EAM & 230 & 12.01 & 153.2 & 66.7 & 71.7 \tabularnewline
\cline{2-8}

Cu$_{}$Fe$_{}$ & B2 & DFT & 264, 262\cite{site:cmu} & 11.88 & 211.4 & 108.7 & $-$52.0 \tabularnewline
         &          & MEAM/EAM & 217 & 12.10 & 161.7 & 105.6 & 45.7 \tabularnewline
\cline{2-8}

\hline

    \end{tabular}
  \end{ruledtabular}
\end{table*}
\endgroup

\subsection{Substitutions}

The formation energy of a substitutional point defect
$E_{\text{f}}^{\text{sub}}$, in the case of the substitution of a
type-X atom of the host with a type-Y atom, is defined by
\begin{equation}
  E_{\text{f}}^{\text{sub}} = E_{\text{tot}}[(N-1)+1] 
  - (N-1) \varepsilon_{\text{X}} - \varepsilon_{\text{Y}}
  \label{eq:Ef_sub}
\end{equation}
where $E_{\text{tot}}[(N-1)+1]$ is the total energy of a system of
$N-1$ host type-X atoms and one type-Y atom that replaced type-X atom
in the original bulk position, $\varepsilon_{\text{X}}$ and
$\varepsilon_{\text{Y}}$ are the total energies per atom for type-X
and type-Y atoms in their ground state bulk structures.
Table~\ref{tab:subst} shows the results of substitutional defect
calculations using the MEAM potentials and the DFT results. In general
the MEAM results qualitatively agree with the DFT results.
In a number of cases of small heat of formation, MEAM indicates a
small heat, but of the incorrect sign.  The most significant error is
for Al in Si where MEAM predicts a large endothermic heat and DFT
predicts a much smaller value, otherwise there is general agreement.

\subsection{Finite temperature tests}
\label{sec:finite_t}

\begin{table}[!tbh]
  \caption{Composition limits of AA~6061 alloy \cite{handbook1998} and
    a model system used to estimate thermal expansion coefficient.
    \label{tab:compos}}
\begin{ruledtabular}
\begin{tabular}{cccc}
  & \multicolumn{2}{c}{Limits} & 
  \multirow{2}{*}{Model [wt.~\%]} \\
  \cline{2-3}
  Element & low [wt.~\%] & high [wt.~\%] & \\
  \hline
  Si   & 0.40 & 0.8  & 0.51  \\
  Mg   & 0.8  & 1.2  & 1.00  \\
  Cu   & 0.15 & 0.40 & 0.30  \\
  Fe   & no   & 0.7  & 0.50  \\
  Mn   & no   & 0.15 & 0.00  \\
  Cr   & 0.04 & 0.35 & 0.00  \\
  Zn   & no   & 0.25 & 0.00  \\
  Ti   & no   & 0.15 & 0.00  \\
\end{tabular}
\end{ruledtabular}
\end{table}

Real life applications of MD potentials require extensive testing at
finite temperatures. Basic finite temperature tests of the potentials,
in accord with recommendations of \citet{lee2010:meam_progr}, revealed
formation of an unknown solid structure when the temperature of fcc Al
crystal was increased to 800~K under zero pressure conditions.  To
prevent formation of this structure, $\beta^{(1)}$ and $t^{(1)}$
parameters of Al were adjusted. Heating of other elements under zero
pressure conditions did not result in forming new structures.

To test a system including all components of the new potential, an
20--100~$^\circ$C average thermal expansion coefficient of a model
system with the composition similar to AA~6061 alloy
(Table~\ref{tab:compos}) was evaluated and compared with experimental
data. Atoms of constituents were placed in the substitutional
positions of a 20x20x20 fcc Al cell. The system was heated from
-200~$^\circ$C to 20~$^\circ$C (and 100~$^\circ$C) over the interval
of 0.1~ns, and then equilibrated at 20~$^\circ$C (and 100~$^\circ$C)
for 1~ns under zero pressure conditions. Table~\ref{tab:finT} shows
the values of 20--100~$^\circ$C average thermal expansion
coefficients. The MEAM result for single crystal Al is in the lower range of
other MD potentials and experiments. Since Al is a dominant element of
the AA~6061 alloy, the thermal expansion coefficient for alloy is
similarly underestimated, possibly also due to imperfections of the
structure of the real material.

\begin{table}[]
\caption{Thermal expansion coefficient of single crystal Al and AA~6061
  alloy between 20$^\circ$C and 100$^\circ$C in the units of
  $\mu$m/m/K.\label{tab:finT}}.
\begin{ruledtabular}
\begin{tabular}{ccccc}
    &\multicolumn{2}{c}{CMD} & \multicolumn{2}{c}{Exp} \\
    \cline{2-5}
    & present &  Ref.~\citep{Chu2011} & Ref.~\citep{wilson1941} & Ref.~\citep{handbook1998}\\
  \hline
  Al fcc & 14.4 & 15-25                   & 25.4 & 23.6\\ 
  AA 6061 & 14.6 &                         &      & 23.6\\
  
\end{tabular}
\end{ruledtabular}
\end{table}

\section{Conclusions}

In this study we developed MEAM potentials for the pair combinations
of aluminum, silicon, magnesium, copper, and iron.  The MEAM formalism
allows any of these potentials to be combined to enable prediction of
multi-component alloy properties. These potentials reproduce a large
body of elemental and binary properties from DFT calculations at the
temperature of 0~K and experimental results. Basic finite temperature
tests of the single element potentials and their alloy combinations
were also performed. With focus to facilitate reproducibility of the
presented results~\cite{site:ase_tests}, and subject to further
testing and improvements, these potentials are one step towards
designing multi-component alloys by simulations.

\section*{Acknowledgment}

The authors are grateful to the Center for Advanced Vehicular Systems
at Mississippi State University for supporting this study.  Computer
time allocation has been provided by the High Performance Computing
Collaboratory (HPC$^2$) at Mississippi State University.
Computational package \textsc{lammps}~\citep{plimpton95:fpa} with
ASE~\citep{bahn2002:oo_scr_int} interface was used to perform MD
simulations. Much appreciated tests of the new MEAM potentials,
including the high temperature simulations of Al that revealed
formation of unknown Al phase at 800~K, were performed by Chandler
Becker and Tanner Hamann at the Metallurgy Division of the Material
Measurement Laboratory, National Institute of Standards and Technology
(NIST). Comparison of ab-initio elastic constants and related
discussion with Hannes Schweiger from Materials Design are also
appreciated. Classical MD potentials from other authors examined in
this study were downloaded from the Interatomic Potentials Repository
Project database \cite{becker2011:atomist_sim}.

Sandia National Laboratories is a multi-program laboratory managed and
operated by Sandia Corporation, a wholly owned subsidiary of Lockheed
Martin Corporation, for the U.S. Department of Energy's National
Nuclear Security Administration under contract DE-AC04-94AL85000.

\appendix

\section{MEAM theory}
\label{sec:meam_theory}

The total energy $E$ of a system of atoms in the
MEAM~\cite{baskes92:modif} is approximated as the sum of the atomic
energies
\begin{equation}
  E = \sum_{i} E_i.
\end{equation}
The energy of atom $i$ consists of the embedding energy and the pair
potential terms:
\begin{equation}
  E_i = F_i\left( \bar\rho_{i} \right) + 
  \frac{1}{2} \sum_{j \neq i}\phi_{ij}\left(r_{ij}\right).
\end{equation}
$F$ is the embedding function, $\bar\rho_{i}$ is the background
electron density at the site of atom $i$, and
$\phi_{ij}\left(r_{ij}\right)$ is the pair potential between atoms $i$
and $j$ separated by a distance $r_{ij}$.  The embedding energy
$F_i\left(\bar\rho_{i}\right)$ represents the energy cost to insert
atom $i$ at a site where the background electron density is
$\bar\rho_{i}$. The embedding energy is given in the form
\begin{equation}
  \label{eq:emb}
  F_i\left(\bar\rho_{i}\right) = 
  \begin{cases}
    A_{i} E_{i}^{0} \bar\rho_{i} \ln \left(\bar\rho_i\right) & 
    \text{if $\bar\rho_{i} \ge 0$},\\
    -A_{i} E_{i}^{0} \bar\rho_{i} &
    \text{if $\bar\rho_{i} < 0$},\\
  \end{cases}
\end{equation}
where the sublimation energy $E_i^0$ and parameter $A_i$ depend on the
element type of atom $i$.  The background electron density
$\bar\rho_i$ is given by
\begin{equation}
  \bar\rho_{i} = \frac{\rho_{i}^{\left( 0 \right)}}{\rho_{i}^0}
  G\left( \Gamma_i \right),
\end{equation}
where
\begin{equation}
  \Gamma_i = \sum_{k=1}^3 t_i^{\left(k\right)}
  \left(
    \frac{\rho_i^{\left(k\right)}}{\rho_i^{\left(0\right)}}
  \right)^2
\end{equation}
and
\begin{equation}
    G\left(\Gamma\right) =
    \begin{cases}
      \sqrt{1 + \Gamma}    & \text{if $\Gamma \geq -1$},\\
      -\sqrt{|1 + \Gamma|} & \text{if $\Gamma < -1$}.\\
    \end{cases}
\end{equation}
The zeroth and higher order densities, $\rho_i^{(0)}$, $\rho_i^{(1)}$,
$\rho_i^{(2)}$, and $\rho_i^{(3)}$ are given in
Eqs.~(\ref{eq:part_den}).  The composition-dependent electron density
scaling $\rho_i^0$ is given by
\begin{equation}
  \rho_i^0 = \rho_{i0}Z_{i0}G\left( \Gamma_i^\text{ref} \right),
\end{equation}
where $\rho_{i0}$ is an element-dependent density scaling, $Z_{i0}$
is the first nearest-neighbor coordination of the reference system, and
$\Gamma_i^\text{ref}$ is given by
\begin{equation}
  \Gamma_i^\text{ref} = \frac{1}{Z_{i0}^2}
  \sum_{k=1}^3 t_i^{\left( k \right)} s_i^{\left( k \right)},
\end{equation}
where $s_i^{(k)}$ is the shape factor that depends on the reference
structure for atom $i$. Shape factors for various structures are
specified in the work of \citet{baskes92:modif}.  The partial
electron densities are given by
\begin{subequations}
  \label{eq:part_den}
\begin{eqnarray}
  \label{eq:part_den_first}
  \rho_i^{\left( 0 \right)} & = &
  \sum_{j \neq i} \rho_j^{a\left( 0 \right)} \left( r_{ij} \right) S_{ij}\\
  \left( \rho_i^{\left( 1 \right)} \right)^2 & = &
  \sum_{\alpha}
  \left[
    \sum_{j \neq i} \rho_j^{a\left( 1 \right)}
    \frac{r_{ij\alpha}}{r_{ij}} S_{ij}
  \right]^2\\
  \left( \rho_i^{\left( 2 \right)} \right)^2 & = &
  \sum_{\alpha, \beta}
  \left[
    \sum_{j \neq i} \rho_j^{a\left( 2 \right)}
    \frac{r_{ij\alpha}r_{ij\beta}}{r_{ij}^2} S_{ij}
  \right]^2\nonumber\\
  & - & \frac{1}{3}
  \left[
    \sum_{j \neq i} \rho_j^{a\left( 2 \right)}
    \left( r_{ij} \right) S_{ij}
  \right]^2
  \\
  \left( \rho_i^{\left( 3 \right)} \right)^2 & = &
  \sum_{\alpha, \beta, \gamma}
  \left[
    \sum_{j \neq i} \rho_j^{a\left( 3 \right)}
    \frac{r_{ij\alpha}r_{ij\beta}r_{ij\gamma}}{r_{ij}^3} S_{ij}
  \right]^2\nonumber\\
  & - & \frac{3}{5} \sum_{\alpha}
  \left[
    \sum_{j \neq i} \rho_j^{a\left( 3 \right)}
    \frac{r_{ij\alpha}}{r_{ij}} S_{ij}
  \right]^2,
  \label{eq:part_den_last}
\end{eqnarray}
\end{subequations}
where $r_{ij\alpha}$ is the $\alpha$ component of the displacement vector
from atom $i$ to atom $j$.  $S_{ij}$ is the screening function between
atoms $i$ and $j$ and is defined in Eqs.~(\ref{eq:scr}).  The atomic
electron densities are computed as
\begin{equation}
  \rho_i^{a\left( k \right)}\left( r_{ij}\right) =
  \rho_{i0} \exp
  \left[
    - \beta_i^{\left( k \right)} \left( \frac{r_{ij}}{r_i^0} - 1 \right)
  \right],
\end{equation}
where $r_i^0$ is the nearest-neighbor distance in the single-element
reference structure and $\beta_i^{\left( k \right)}$ is
element-dependent parameter.
Finally, the average weighting factors are given by
\begin{equation}
  t_i^{\left( k \right)} = 
  \frac{
    \sum_{j \neq i} t_{0, j}^{\left( k \right)} \rho_j^{a\left( 0 \right)} S_{ij}
  }
  {
    \sum_{j \neq i} \left(t_{0, j}^{\left( k \right)}\right)^2 \rho_j^{a\left( 0 \right)} S_{ij}
  }
  ,
\end{equation}
where $t_{0,j}^{\left( k \right)}$ is an element-dependent parameter.

The pair potential is given by
\begin{align}
  \label{eq:pair}
  \phi_{ij}\left(r_{ij}\right) &= \bar\phi_{ij}\left(r_{ij}\right)
  S_{ij}\\
  \begin{split}
  \bar\phi_{ij}\left(r_{ij}\right) &= \frac{1}{Z_{ij}}
  \left[ 2E_{ij}^u \left( r_{ij} \right) 
    -F_i\left(\frac{Z_{ij}}{Z_i} \rho_j^{a(0)} 
  \left( r_{ij} \right) \right) \right. \\
  &\quad \left. -F_j\left(\frac{Z_{ij}}{Z_j} \rho_j^{a(0)} 
  \left( r_{ij} \right) \right) \right] \label{eq:rhohat}
\end{split} \\
  E_{ij}^u\left( r_{ij} \right) 
  &= -E_{ij}\left( 1 + a_{ij}^*\left(r_{ij}\right)\right)
  e^{-a_{ij}^{*}\left(r_{ij}\right)}\\
  a_{ij}^{*} &= \alpha_{ij} \left( \frac{r_{ij}}{r_{ij}^0} - 1 \right),
\end{align}
where $E_{ij}$, $\alpha_{ij}$ and $r_{ij}^0$ are element-dependent
parameters and $Z_{ij}$ depends upon the structure of the reference
system. The background densities $\hat\rho_i(r_{ij})$ in
Eq.~(\ref{eq:rhohat}) are the densities for the reference structure
computed with interatomic spacing $r_{ij}$.

The screening function $S_{ij}$ is designed so that $S_{ij} = 1$ if
atoms $i$ and $j$ are unscreened and within the cutoff radius $r_c$,
and $S_{ij} = 0$ if they are completely screened or outside the cutoff
radius. It varies smoothly between 0 and 1 for partial screening. The
total screening function is the product of a radial cutoff function
and three body terms involving all other atoms in the system:
\begin{subequations}
  \label{eq:scr}
  \begin{align}
    \label{eq:scr_first}
    S_{ij} &= \bar S_{ij} f_c \left( \frac{r_c - r_{ij}}{\Delta r} \right)\\
    \bar S_{ij} &= \prod_{k\ne i,j}S_{ikj}\\
    S_{ikj} &= f_c \left(\frac{C_{ikj} - C_{\text{min},ikj}}
      {C_{\text{max},ikj} - C_{\text{min},ikj}} \right)\\
    C_{ikj} &= 1 + 2 \frac{r_{ij}^2 r_{ik}^2 + r_{ij}^2 r_{jk}^2 
      - r_{ij}^4}{r_{ij}^4 - \left( r_{ik}^2 - r_{jk}^2 \right)^2}\\
    f_c\left(x\right) &=
    \begin{cases}
      1 & x \geq 1\\
      \left[ 1 - \left( 1 - x )^4 \right) \right]^2 & 0<x<1\\
      0 & x \leq 0\\
    \end{cases}
    \label{eq:scr_last}
  \end{align}
\end{subequations}
Note that $C_{\text{min}}$ and $C_{\text{max}}$ can be defined
separately for each $i$-$j$-$k$ triplet, based on their element
types. The parameter $\Delta r$ controls the distance over which the
radial cutoff is smoothed from 1 to 0 near $r=r_\text{c}$.

\section{Equilibrium lattice parameter and bulk modulus}

MEAM postulates the Rose universal equation of
state\cite{rose84:univ_eos}
\begin{equation}
  E_\text{R}(a^*)
  =
  -E_\text{c}
  \left(1 + a^* + \delta \frac{\alpha a^{*3}}{\alpha+a^*}\right)
  e^{-a^*}
  \label{eq:rose1}
\end{equation}
for the reference structure of each single element and for each
element pair. The $a^*$, scaled distance from the equilibrium nearest
neighbor position $r_0$, is
\begin{equation}
a^*
=
\alpha(r/r_0-1).
\end{equation}
Two $\delta$ parameters may be specified for each element/pair:
$\delta_r$ for negative, and $\delta_a$ for positive $a^*$. Then
\begin{equation}
  \delta=\begin{cases}
    \delta_r & \text{ for }a^* < 0\\
    \delta_a & \text{ for }a^* \geq 0.
  \end{cases}
\end{equation}
The MEAM potential parameter $\alpha$ is related to
the equilibrium atomic volume $\Omega_0$, the bulk modulus $B_0$, and
the cohesive energy of the reference structure $E_\text{c}$ as follows
\begin{equation}
\alpha
=
\sqrt{\frac{9B_0\Omega_0}{E_\text{c}}}.
\label{eq:rose4}
\end{equation}

The DFT equilibrium energies and bulk moduli were obtained by fitting
energy-volume dependence to Murnaghan equation of
state~\cite{murnaghan1944:compress}
\begin{eqnarray}
  E\left(V\right) &=&E(V_0)\\
  &+&\frac {B_0 V} {B_0' (B_0' - 1)}
  \left [
    B_0' \left(1 - \frac{V_0}{V}\right)
    + \left(\frac{V_0}{V}\right) ^{B_0'}
    - 1
  \right].\nonumber
  \label{eq:murn}
\end{eqnarray}

\section{Trigonal and tetragonal shear modulus}

For small deformations of a cubic crystal, the change of energy density due
to straining is
\begin{eqnarray}
  \triangle E_V &=&
  \frac{1}{2} C_{11} \left(\epsilon_1^2 + \epsilon_2^2 + \epsilon_3^2\right)
  + C_{12} \left(\epsilon_1\epsilon_2 + \epsilon_2\epsilon_3 + \epsilon_3\epsilon_1\right) \nonumber\\
  &+& \frac{1}{2} C_{44} \left( \epsilon_4^2 + \epsilon_5^2 + \epsilon_6^2 \right)
  + O(\epsilon_i^3),
  \label{eq:cub}
\end{eqnarray}
where $\epsilon_{i}$ are strains in modified Voigt notation.

The trigonal shear modulus $C_{44}$ was determined from rhombohedral
deformation given by $\epsilon_1 = \epsilon_2 = \epsilon_3 = 0$ and
$\epsilon_4 = \epsilon_5 = \epsilon_6 = \delta$ in
\ref{eq:cub}, leading to
\begin{equation}
  \triangle E_V(\delta) = \frac{3}{2} C_{44} \delta^2 + O(\delta^3).
  \label{eq:trig}
\end{equation}

The tetragonal shear modulus $(C_{11}-C_{12})/2$ was determined
from the deformation given by $\epsilon_1 = \delta, \epsilon_2 = \frac{1}{1+\delta}-1$ in
\ref{eq:cub}, leading to
\begin{equation}
  \triangle E_V (\delta) = \left(C_{11} - C_{12}\right) \delta^2 + O(\delta^3).
  \label{eq:tetr}
\end{equation}

The \ref{eq:trig} and \ref{eq:tetr} were used to estimate tetragonal
and trigonal shear moduli.

\end{document}